\documentclass[12pt]{article}
\pdfoutput=1
\usepackage{amsmath,amsfonts,amsthm,amssymb,slashed,graphicx}
\usepackage{xcolor}
\usepackage{graphicx}
\usepackage{epstopdf}
\usepackage{array,booktabs}
\usepackage{hyperref}
\usepackage{tabulary}
\usepackage{multirow}
\usepackage{array,arydshln}
\usepackage{bbm} % for \mathbbm{1}

\hypersetup{linktoc = all}  %hyperref settings
\hypersetup{pdfborderstyle={/S/U/W 0.5}}

\numberwithin{equation}{section}
\def\CA{{\cal A}}

\def\CJ{{\cal J}}
\def\CL{{\cal L}}
\def\CN{{\cal N}}\def\CO{{\cal O}}
\def\CQ{{\cal Q}}\def\CR{{\cal R}}

\def\CW{{\cal W}}

\def\a{\alpha}\def\g{\gamma}
\def\d{\delta}\def\e{\epsilon}
\def\z{\zeta}\def\th{\theta}
\def\k{\kappa}\def\l{\lambda}
\def\m{\mu}
\def\r{\rho}\def\s{\sigma}
\def\t{\tau}
\def\w{\omega}\def\G{\Gamma}
\def\D{\Delta}\def\L{\Lambda}
\def\S{\Sigma}

\newcommand{\address}[1]{\vbox{\center\em#1}}
\renewcommand{\title}[1]{\vbox{\center\LARGE{#1}}\vspace{5mm}}

\newcommand{\beq}{\begin{equation}}
\newcommand{\eeq}{\end{equation}}

\newcommand{\bra}[1]{\langle #1|}

\newcommand{\rf}[1]{\eqref{#1}}
\declareslashed{}{/}{-.08}{0}{V} % place slash in Vslash.

\makeatletter

\newcommand*{\letterdef@}{}
\newcommand*{\letterdef}[3]{%
  \def\letterdef@##1{\expandafter\newcommand\csname #1\endcsname{#2{##1}}}%
  \@tfor\@tempa :=#3\do{\expandafter\letterdef@\expandafter{\@tempa}}}
\makeatother
\letterdef{b#1} {\mathbb} {CHPRZ}    % \bX = \mathbb{X} for X in {CHPRZ}.
\letterdef{c#1} {\mathcal}{ABCDEFGHIJKLMNOPQRSTUVWXYZ} % \cX = \mathcal{X}
\letterdef{rm#1}{\mathrm} {dDmM} % \rmX = \mathrm{X} for X in {dDmM}
\letterdef{f#1}{\mathfrak}{Btqr}

\begin{document}
\bibliographystyle{utphys}

%%%%%%%%%%%%%%%%%%%%%%%%%%%%%%%%%%%%%%%%%%%%%%%%%%%%%%%%%%%%%%%%%%
%%%%%%%%%%%%%%%%%%%%%%%%%%%%%%%%%%%%%%%%%%%%%%%%%%%%%%%%%%%%%%%%%%

\begin{titlepage}
 \vspace{10mm}

  \hfill {\tt DAMTP-2012-68}\\
  \vspace{25mm}

  \centerline{\LARGE{\scalebox{0.9}{Exact K\"ahler Potential  from Gauge Theory  and Mirror Symmetry}}}
  \vspace{10mm}
  \begin{center}
    \renewcommand{\thefootnote}{$\alph{footnote}$}
    Jaume Gomis\footnote{\href{mailto:jgomis@perimeterinstitute.ca}{\tt jgomis@perimeterinstitute.ca}}
    and Sungjay Lee\footnote{\href{mailto:S.Lee@damtp.cam.ac.uk}{\tt S.Lee@damtp.cam.ac.uk}}
    \vspace{15mm}

    \address{${}^{a}$Perimeter Institute for Theoretical Physics,\\ Waterloo, Ontario, N2L 2Y5, Canada}
    \address{${}^{b}$DAMTP, Centre for Mathematical Sciences,\\ Cambridge University, Cambridge CB3 0WA,
    United Kingdom}
  \end{center}
 \vspace{20mm}

  \abstract{\medskip\medskip\normalsize{
  \noindent We prove a recent conjecture that the partition function of $\cN=(2, 2)$ gauge theories on the two-sphere which flow to Calabi-Yau sigma models in the infrared computes the exact K\"ahler potential on the quantum K\"ahler moduli space of the corresponding Calabi-Yau.  This establishes the two-sphere partition function
  as a new method of computation of  worldsheet instantons and    Gromov-Witten invariants. We  also calculate  the exact two-sphere partition function for $\cN=(2,2)$
  Landau-Ginzburg models with an arbitrary twisted superpotential $W$.  These results are used to demonstrate that arbitrary abelian gauge theories and their   associated  mirror Landau-Ginzburg models have identical two-sphere partition functions. We further show that    the      partition function of non-abelian gauge theories  can be   rewritten as the partition function of  mirror Landau-Ginzburg models.
  }}
  \vspace{10mm}

  \noindent
  %\today
  \vfill

\end{titlepage}

\renewcommand{\thefootnote}{\arabic{footnote}}
\setcounter{footnote}{0}

\tableofcontents

%%%%%%%%%%%%%%%%%%%%%%%%%%%%%%%%%%%%%%%%%%%%%%%%%%%%%%%%%%%%%%%%%%
%%%%%%%%%%%%%%%%%%%%%%%%%%%%%%%%%%%%%%%%%%%%%%%%%%%%%%%%%%%%%%%%%%

\section{Introduction}
\label{sec:intro}

Two dimensional gauge theories  -- known as gauged linear sigma models (GLSM) --
provide a variety of insights into the dynamics of nonlinear sigma models \cite{Witten:1993yc}.
These two dimensional field theories -- apart from mimicking  ubiquitous phenomena  in
four dimensional gauge theories -- play a central role in string  theory.
Specially, nonlinear sigma models on Calabi-Yau threefold target spaces, which give rise to
a very rich set of vacua of string theory. The  study of Calabi-Yau sigma models is at
the genesis of several  remarkable discoveries --  including mirror symmetry \cite{Dixon:1987bg,Lerche:1989uy}-- and
has emerged as a source of inspiration for both physicists and mathematicians.

Physical observables in Calabi-Yau sigma models can receive non-perturbative corrections.
These are generated by worldsheet instantons, which correspond to
holomorphic maps from the worldsheet to the Calabi-Yau target space.
Worldsheet instanton corrections to the point particle approximation correct the
K\"ahler moduli space of the Calabi-Yau manifold into the so called ``quantum K\"ahler moduli space",
from which effective Yukawa couplings among  other observables can be computed.
The problem of summing over all worldsheet instantons defines an interesting class of topological invariants,
known as the Gromow-Witten invariants \cite{MR809718,Witten:1988xj}. One of the central themes in this research area
is to develop methods to compute these invariants.

Recently --  with Doroud and Le Floch --  we have computed \cite{Doroud:2012xw}
the exact partition function of $\cN=(2,2)$ gauge theories on $S^2$ (see also \cite{Benini:2012ui}).
The partition function admits two alternative descriptions, either as an integral over the Coulomb branch
or as sum over vortex and anti-vortex configurations on the Higgs vacua of the theory.
Each representation yields complementary insights into the dynamics of these gauge theories.
These results offer a new window into the exact dynamics of nonlinear sigma models on K\"ahler manifolds.
Specifically  on Calabi-Yau manifolds, since  two dimensional $\cN=(2,2)$ gauge theories  which flow to an
$\cN=(2,2)$ superconformal field theory  in the infrared provide an elegant  framework to study
Calabi-Yau sigma models \cite{Witten:1993yc} (see also \cite{Morrison:1994fr}).

In \cite{Doroud:2012xw} these exact results were applied to the study of worldsheet instantons in
Calabi-Yau sigma models. The representation of the $S^2$ partition function as a sum over vortices
at the north pole and anti-vortices at the south pole was used  to explicitly confirm that the
non-perturbative worldsheet instanton  corrections  to the K\"ahler moduli space de-singularize
the quantum dynamics across the topology-changing flop transition
\cite{Witten:1993yc,Aspinwall:1993yb}.\footnote{The flop transition was shown to
correspond to crossing symmetry in a dual  Toda CFT correlator,
which computes the $S^2$ partition function for a class of two dimensional theories.}
We further speculated that the exact partition function of  $\cN=(2,2)$ gauge theories on $S^2$
may provide a novel approach to the  computation of worldsheet instantons in Calabi-Yau sigma models.

More recently, these speculations became a conjecture in an interesting  paper by Jockers {\it et al.}
\cite{Jockers:2012dk}. These authors conjectured that the exact $S^2$ partition function $Z$
\cite{Doroud:2012xw,Benini:2012ui} of an $\cN=(2,2)$ gauge theory flowing to a sigma model on a
Calabi-Yau computes the exact K\"ahler potential $\cK$ on the quantum K\"ahler moduli space $\cM$
of the Calabi-Yau
\beq
Z(\tau_a,\bar \tau_a)=e^{-\cK(\tau_a,\bar \tau_a)}\,.
\label{kahler}
\eeq
Here, $\tau_a$ are coordinates in $\cM$  parametrizing the K\"ahler moduli of the Calabi-Yau,
which correspond to the complexified Fayet-Iliopoulos parameters in the GLSM.
Evidence for this conjecture was presented in \cite{Jockers:2012dk} by extracting
the Gromov-Witten invariants from the exact K\"ahler potential $\cK$ via \rf{kahler} and matching them
with those computed in the literature using different methods.
Remarkably, the partition function on $S^2$ \cite{Doroud:2012xw,Benini:2012ui} was also used
in \cite{Jockers:2012dk} to predict new Gromov-Witten invariants, which pass all consistency checks.

In this paper  we provide  two alternative proofs  of the conjecture \rf{kahler}.
One derivation uses the gauge theory approach to Calabi-Yau sigma models provided by GLSM's.
We demonstrate that path integral on the {\it squashed} two-sphere in the limit in which
the two-sphere is infinitely squashed describes a very specific overlap of ground states of the
infrared conformal field theory, which is known from the work of  Cecotti and Vafa \cite{Cecotti:1991me}
to compute the exact K\"ahler potential $\cK$.
By explicitly proving that the partition function on the squashed two-sphere does not depend on the
squashing parameter, we arrive to \rf{kahler}. This chain of reasoning ends up relating the partition function on the round two-sphere
with the exact K\"ahler potential in the quantum K\"ahler moduli space of the Calabi-Yau
that emerges in the infrared.

We also compute the exact two-sphere  partition function of $\cN=(2,2)$ Landau-Ginzburg models with
an arbitrary twisted superpotential $W$. The partition function takes the simple form
\begin{align}
  Z= \int dY d\overline Ye^{ - 4\pi i r W(Y) - 4\pi i r \overline{W}(\overline Y)}\,,
  \label{twistpartitionaa}
\end{align}
where $r$ is the radius of the two-sphere. Landau-Ginzburg models capture the dynamics of   Calabi-Yau non-linear sigma models   in certain domains
of the  K\"ahler moduli space of a Calabi-Yau as well as serving as the mirror description of Calabi-Yau sigma
models. We  study  the two-sphere partition function  as a function of the space of marginal deformations of the
underlying superconformal field theory -- which span the K\"ahler moduli space $\cM$ -- and  obtain a
second derivation  of the conjecture \rf{kahler}, now from the Landau-Ginzburg approach to
Calabi-Yau sigma models.

We investigate mirror symmetry for sigma models on K\"ahler manifolds -- including Calabi-Yau manifolds --
from the viewpoint of the two-sphere partition function.
This uses the results for the partition function of GLSM's found in \cite{Doroud:2012xw,Benini:2012ui}
and the result  \rf{twistpartitionaa} derived in this paper for Landau-Ginzburg models.
We show that the $S^2$ partition function of the Landau-Ginzburg models put forward
by Hori and Vafa \cite{Hori:2000kt} exactly reproduces the $S^2$ partition function for
the mirror abelian GLSM's, which describe toric varieties and complete intersections in toric varieties.
For non-abelian GLSM's we use the exact  results on the two-sphere \cite{Doroud:2012xw,Benini:2012ui}
to rewrite the gauge theory partition function in Landau-Ginzburg form and
prove a conjecture by Hori and Vafa   \cite{Hori:2000kt} for the mirror Landau-Ginzburg description
of these non-abelian GLSM's.

The plan of the rest of the paper is as follows. In section \ref{sec:SUSY} we construct the Lagrangian and
supersymmetry transformations of $\cN=(2,2)$ theories on the squashed two-sphere.
In section \ref{sec:overlap} we provide a proof of the conjecture \rf{kahler} in two steps.
We  first prove that the partition function on the squashed two-sphere is independent of the
squashing parameter, while relegating many of the technical details of this proof to the Appendix.
We then  show that the path integral on the squashed two-sphere in the limit where
the sphere is infinitely squashed provides a path integral representation of
the ground state overlap which is known to compute the K\"ahler potential $\cK$.
In section \ref{sec:mirror} we construct  the  Lagrangian and supersymmetry transformations of $\cN=(2,2)$
Landau-Ginzburg models with an arbitrary twisted superpotential $W$.
We then use these results to  give an alternative proof  of the conjecture  \rf{kahler} as well as
to establish the exact equivalence of the two-sphere partition function of GLSM's
and mirror Landau-Ginzburg models.

 %%%%%%%%%%%%%%%%%%%%%%%%%%%%%%%%%%%%%%%%%%%%%%%%%%%%%%%%%%%%%%%%%%
%%%%%%%%%%%%%%%%%%%%%%%%%%%%%%%%%%%%%%%%%%%%%%%%%%%%%%%%%%%%%%%%%%

\section{Supersymmetric Theories on Squashed Two-Sphere}
\label{sec:SUSY}

The partition function  of $\cN=(2,2)$ gauge theories on $S^2$ was computed in
\cite{Doroud:2012xw,Benini:2012ui}.\footnote{We follow the notation and conventions in \cite{Doroud:2012xw},
which can be consulted for more details.} These theories are invariant under the $SU(2|1)$ algebra, which is
the  $\cN=(2,2)$  supersymmetry algebra on $S^2$. In order for a field theory to be supersymmetric on $S^2$,
the theory must admit a $U(1)$ $R$-symmetry, as the associated conserved charge $R$ appears in the
anticommutator of supercharges in $SU(2|1)$.

In this section we construct the supersymmetry transformations and supersymmetric action for $\cN=(2,2)$ gauge theories
on the squashed two-sphere $S^2_b$. As we shall see, the existence of a $U(1)$ $R$-symmetry plays an important role in the construction of these theories.

As we shall see, the study of the partition function of $\cN=(2,2)$ gauge theories
on the squashed two-sphere provides a path towards proving the conjecture \rf{kahler}. In section \ref{sec:LGG}
an alternative approach to the derivation of  \rf{kahler}   is presented.

%%%%%%%%%%%%%%%%%%%%%%%%%%%%%%%%%%%%%%%%%%%%%%%%%%%%%%%%%%%%%%%%%%

\subsection{Superalgebra and Killing Spinors on  Squashed Two-Sphere}

Let us consider deforming the round sphere to the squashed two-sphere $S^2_b$
while preserving a $U(1)\subset SU(2)$ isometry. The squashed two-sphere
can be described by an embedding equation in $\mathbb{R}^3$
\begin{align}
  \frac{x_1^2 + x_2^2}{l^2} + \frac{x_3^2}{\tilde l^2}=1\ ,
\end{align}
which depends on the dimensionless squashing parameter $b=l/\tilde l$.
The metric on the squashed two-sphere is
\begin{align}
  ds^2 = f^2(\th) d\th^2 + l^2 \sin^2\th d\varphi^2 \,,
  \label{metricsq}
\end{align}
where $f^2(\th) = \tilde l^{2} \sin^2 \th + l^2 \cos^2 \th$.
The vielbein in a patch around the equator excluding the north and south poles of $S^2_b$ is chosen as
\begin{align}
  e^{\hat 1} = f(\th) d\th \ , \qquad e^{\hat 2} = l \sin\th d\varphi\,.
  \label{vielb}
\end{align}
In this patch, the spin connection on $S^2_b$ is given by
\begin{align}
 \w\equiv \w_{\hat 1\hat 2} =  - \frac{l\cos\th}{f(\th)}d\varphi\,.
 \label{spin}
\end{align}
Later we consider vielbein and spin connection which are smooth in a patch near the north and the south poles
of $S^2_b$ to analyze the physics near the poles.

Since   squashing the $S^2$ breaks the $SU(2)$ symmetry down to $U(1)$,  the supercharges on $S^2$
that do not generate the $U(1)$ symmetry of $S^2_b$ must be broken on the squashed two-sphere.
This implies that the supersymmetry algebra on the squashed two-sphere   is an $SU(1|1)$ subalgebra of $SU(2|1)$
\begin{align}
  \cQ^2=J+\frac{R}{2} \ ,  \qquad \Big[J+\frac{R}{2},\cQ\Big]=0\,,
  \label{SUSSalg}
\end{align}
where $J$ corresponds to  the $U(1)$ subgroup of the $SU(2)$ symmetry preserving $S^2_b$ and
$R$ is the $U(1)$ $R$-symmetry generator in $SU(2|1)$.
We note that this is precisely the supersymmetry algebra generated by the supercharge used in
\cite{Doroud:2012xw}
\begin{align}
  \cQ= S_1+Q_2\,
  \label{choice}
\end{align}
to localize the path integral of $\cN=(2,2)$ gauge theories on $S^2$.

While bosonic space-time transformations are parametrized by Killing vectors,
Killing spinors parametrize supersymmetry transformations.
On the round two-sphere $S^2$ of radius $l$, the supersymmetry parameters
$\epsilon$ and $\bar{\epsilon}$ corresponding to the supercharge  $\cQ$ in \rf{choice}
are conformal Killing spinors, which can be taken to satisfy
\begin{align}
  \nabla_{i}\epsilon = +\frac{1}{2l}\gamma_{i}\gamma^{3}\epsilon\,,\qquad
  \nabla_{i}\bar{\epsilon} = -\frac{1}{2l}\gamma_{i}\gamma^{ 3}\bar{\epsilon}\,.
  \label{killing}
\end{align}
In the patch near the equator defined by the vielbein \rf{vielb},
they are explicitly given by \cite{Doroud:2012xw}
\begin{align}
  \e = e^{-\frac{i}{2} \g^{\hat 2} \th } e^{+\frac i2 \varphi} \e_\circ \ , \ & \text{ with } \
  \g^3 \e_\circ = + \e_\circ \ ,
  \nonumber \\
  \bar \e  = e^{+\frac{i}{2} \g^{\hat 2} \th } e^{-\frac i2  \varphi} \bar \e_\circ\ ,
  \ & \text{ with } \  \g^3 \bar \e_\circ = - \bar \e_\circ \ .
  \label{killspinQ}
\end{align}

Killing spinors   generating the $SU(1|1)$ supersymmetry transformations on the squashed
two-sphere $S^2_b$ can be found by turning on a   background gauge field $V_i$ for the
$U(1)$ $R$-symmetry, just as  on the squashed three-sphere \cite{Hama:2011ea}.
The Killing spinors on $S^2_b$   satisfy a generalized Killing spinor equation\footnote{This is a
particularly convenient  choice of basis of spinors and gauge which solve the generalized conformal Killing spinor equation
$D_i\eta =\gamma_i \tilde \eta$.}
\begin{align}
  D_i \e = \frac{1}{2f(\th)} \g_i \g^{3} \e \,, \qquad\qquad
  D_i \bar \e = - \frac{1}{2f(\th)} \g_i \g^{3} \bar \e \,,
  \label{general}
\end{align}
where the covariant derivative $D_i$  includes the background gauge connection for the $U(1)$ $R$-symmetry
\begin{align}
  D_i \e = \left( \nabla_i - i V_i \right) \e\ , \qquad
  D_i \bar \e = \left(  \nabla_i + i V_i \right) \bar \e\,,
\end{align}
and we have taken into account that the $R$-charge of $\epsilon$   is $1$ and
that of $\bar\epsilon$   is $-1$. With the choice of connection
\begin{align}
  V = \frac12 \left( 1 - \frac{l}{f(\th)} \right) d\varphi\,,
  \label{backgrounda}
\end{align}
the Killing spinors on $S^2_b$ in the patch near the equator of the squashed two-sphere
are given by \rf{killspinQ}. These generate the $SU(1|1)$ transformations on the squashed two-sphere.
We collect  useful identities obeyed by the Killing spinors on the squashed two-sphere which are
frequently used throughout the paper in the Appendix.

We note that   the background connection for the $U(1)$ $R$-symmetry \rf{backgrounda}
is topologically trivial and is defined everywhere on the squashed two-sphere,
unlike the vielbein and spin connection, that need to be defined in coordinate patches.

%%%%%%%%%%%%%%%%%%%%%%%%%%%%%%%%%%%%%%%%%%%%%%%%%%%%%%%%%%%%%%%%%%

\subsection{Supersymmetric Lagrangian on Squashed Two-Sphere}

The  supersymmetry transformations are similar to those on the round two-sphere
$S^2$ once the background gauge field $V_i$ is coupled to the various fields according to their
$U(1)$ $R$-charges and we take into account the generalized Killing spinor equation \rf{general}.
As explained in \cite{Doroud:2012xw}, the supersymmetry transformations are   completely determined
by demanding that  the supersymmetry  transformations of the theory   in flat space  are covariant
under Weyl transformations (see Appendix B of \cite{Doroud:2012xw} for more details).

Let us begin with the supersymmetry transformations and Lagrangian for the
vector multiplet. The transformation laws generated by the supercharge $\cQ$ are given by
\begin{align}
  \d \l = & (i V_m \g^m    - {\rm D}) \e
  \nonumber \\
  \d \bar \l = & ( i \bar V_m\g^m   + {\rm D}) \bar \e    \nonumber\\
  \d A_i = & - \frac i2 \Big( \bar \e \g_i \l - \bar \l \g_i \e \Big)  \nonumber
  \\
  \d \s_1 = &   \frac12 \Big( \bar \e \l - \bar \l \e \Big)
  \nonumber \\
  \d \s_2 = & - \frac i2 \Big( \bar \e \g_3 \l - \bar \l \g_3 \e \Big)
    \nonumber \\
  \d D = & - \frac i2 \bar \e \slashed{D}  \l - \frac i2 \big[ \s_1, \bar \e \l \big]
  - \frac12 \big[ \s_2 , \bar \e \g^3 \l \big]+ \frac i2 \e \slashed{D}  \bar \l
  - \frac i2 \big[ \s_1 , \bar \l \e \big] - \frac12 \big[ \s_2 , \bar \l \g^3 \e \big] \,,
  \label{SUSYvec}
\end{align}
with $\epsilon$ and $\bar\epsilon$ the Killing spinors on $S^2_b$ and $V_m$ and ${\bar V}_m$ defined by
\begin{align}
  V_m = & \Big( + D_1 \s_1 + D_2 \s_2 ,\  + D_2 \s_1 - D_1 \s_2 ,
  \ F_{\hat 1\hat 2} + i [\s_1,\s_2] + \frac{1}{f(\theta)}\s_1 \Big)\ ,
  \nonumber \\
  \bar V_m = & \Big( - D_1 \s_1 + D_2 \s_2 ,\ - D_2 \s_1 - D_1 \s_2 ,
  \ F_{\hat 1\hat 2} - i [\s_1,\s_2] + \frac{1}{f(\theta)}\s_1 \Big)\ .
\end{align}
Note that in all formulas the covariant derivative $D_i$ includes the gauge,
local Lorentz and $U(1)$ $R$-symmetry connections.
These supersymmetry transformations close off-shell and realize $SU(1|1)$ on the vector multiplet fields.

The Lagrangian for a vector multiplet takes the following form
\begin{align}
  \CL_\text{v.m.} & =  \frac{1}{2g^2} \text{Tr}
  \bigg[ \left(F_{\hat 1\hat 2} + \frac{\s_1}{f(\theta)}\right)^2 + (D_i \s_1)^2 + (D_i \s_2)^2 -[\s_1,\s_2]^2+  {\rm D}^2
  \nonumber \\ & \qquad \qquad \ \
  + i \bar \lambda \gamma^i D_i   \lambda
  + i \bar \l [ \s_1 , \l ] +  \bar \l \g^3 [ \s_2 , \l ]  \bigg]\ ,
  \label{Lagvec}
\end{align}
which can be shown to be invariant under the supersymmetry transformations (\ref{SUSYvec}).
For each $U(1)$ factor in the gauge group $G$, a supersymmetric Fayet-Iliopoulos (FI) ${\rm D}$-term
can be added
\begin{align}
  \CL_\text{FI} = - i \xi \text{Tr}\left[ {\rm D} - \frac{\s_2}{f(\theta)} \right]\ ,
\end{align}
as well as a two-dimensional theta angle
\begin{align}
  \CL_\text{top} = - i \frac{\vartheta}{2\pi} \text{Tr} F \ .
\end{align}

We now in turn consider a chiral multiplet charged in some representation of the gauge group $G$.
The supersymmetry transformations for a chiral multiplet of $R$-charge $-q$ are given by
\begin{align}
  \d \phi = & \bar \e \psi \   \nonumber \\
  \d \bar \phi = & \e \bar \psi \ , \nonumber \\
  \d \psi = & + i \g^i \e D_i \phi + i\e \s_1 \phi + \g^3 \e \s_2 \phi + i \frac{q}{2f(\theta)}
  \g_3 \e \phi + \bar \e F\  \nonumber \\
  \d \bar \psi = & - i \bar \e \g^i D_i \bar \phi + i \bar \e \bar \phi \s_1
  + \bar \e \g^3 \bar \phi \s_2 + i \frac{q}{2f(\theta)} \bar \e \g_3 \bar \phi + \e \bar F\
  \nonumber \\
  \d F  = & \e \Big( i \g^i D_i \psi  - i \s_1 \psi + \g^3 \s_2 \psi - i \l \phi \Big)
  - i \frac{q}{2} \psi \g^i D_i \e \   \nonumber \\
  \d \bar F  = & \bar \e \Big( i \g^i D_i \bar\psi - i\bar\psi \s_1 - \g^3 \bar \psi \s_2
  + i \bar \phi \l \Big) - i \frac{q}{2} \bar \psi \g^i D_i \bar \e\,.
  \label{SUSYmatter}
\end{align}
One can explicitly show that the supersymmetry algebra also closes off-shell for
the chiral multiplet.

The $SU(1|1)$ invariant Lagrangian for a  chiral multiplet of $R$-charge $-q$
is
\begin{align}
  \CL_\text{c.m.} & =  D_i\bar \phi D^i \phi + \bar \phi \left[ \s_1^2  +
  \s_2^2 + i \frac{q-1}{f(\theta)} \s_2  - \frac{q^2}{4f(\theta)^2} + \frac{q}{4} \CR  \right] \phi
  + \bar F F + i \bar \phi {\rm D} \phi
  \nonumber \\ & \ \ \
  - i \bar \psi \g^i D_i \psi + \bar \psi \left[ i\s_1 - \left( \s_2 + i\frac{q}{2f(\theta)} \right) \g^3 \right]\psi
  + i\bar \psi \l \phi - i \bar \phi \bar \l \psi \ ,
  \label{Lagmatter}
\end{align}
where $\CR$ denotes the curvature scalar of the squashed two-sphere.
A supersymmetric superpotential $\CW$ term  for chiral multiplets
\begin{align}
  \CL_\CW = F_\CW + {\bar F}_\CW\ ,
  \label{superp}
\end{align}
can be added provided that the superpotential  carries $R$-charge $-2$, i.e., $q_\CW=2$.
The supersymmetric superpotential couplings \rf{superp} are precisely those of the theory in flat space, even though the action
of a supersymmetric theory on curved space is generically corrected by $1/r$ terms \cite{Festuccia:2011ws}.

When the theory has a flavour symmetry group $G_F$, we can turn on a supersymmetric background expectation value
for a non-dynamical $G_F$ vector multiplet. This introduces twisted masses $m$   for the chiral multiplets.
As we shall prove, the twisted masses $m$ and $R$-charges $q$ appear in the expression we derive
for the partition function in the  holomorphic combination
\begin{align}
  M=m+i{q\over 2l}\,.
  \label{holol}
\end{align}
These  take values in the Cartan subalgebra of $G_F$.

 It is also possible to introduce    twisted chiral multiplets    preserving  supersymmetry on the squashed-sphere.
 Even though twisted chiral multiplets must be neutral under conventional vector multiplets, these multiplets play an important
 role in mirror symmetry and we will consider them in detail in section \ref{sec:mirror}. As we shall see, the supersymmetrized twisted superpotential $W$
 couplings on the squashed two-sphere are modified in comparison to the theory in flat space. These modifications   have interesting physical implications.

%%%%%%%%%%%%%%%%%%%%%%%%%%%%%%%%%%%%%%%%%%%%%%%%%%%%%%%%%%%%%%%%%%
%%%%%%%%%%%%%%%%%%%%%%%%%%%%%%%%%%%%%%%%%%%%%%%%%%%%%%%%%%%%%%%%%%

\section{A Proof of the Conjecture}
\label{sec:overlap}

The main conjecture that we want to prove states    that whenever a
Calabi-Yau manifold  admits  a Gauged Linear Sigma Model (GLSM) description,
then the  exact partition function of the GLSM on the round two-sphere \cite{Doroud:2012xw,Benini:2012ui}
determines the exact K\"ahler potential in the quantum K\"ahler moduli space, including all
worldsheet instanton corrections. The explicit conjecture is
\begin{align}
  Z\left( \t_a,{ \bar \t}_a \right)
  = e^{- \cK\left(\t_a, \bar \t_a \right)}\,.
\end{align}
This conjecture  \cite{Jockers:2012dk} successfully passes several nontrivial tests and
has been used to predict new  Gromov-Witten invariants for a  Calabi-Yau manifold for
which no  mirror Calabi-Yau is known.

Before delving into a proof of this conjecture,  we will make a few preparatory remarks.
In the GLSM approach to  nonlinear sigma models on Calabi-Yau manifolds,
the moduli of the  Calabi-Yau map to parameters in the associated GLSM.
The complex structure deformations of the Calabi-Yau manifold appear as parameters in the superpotential $\cW$.
These parameters   can be interpreted  as background expectation values for chiral multiplets.
The (complexified) K\"ahler parameters of the Calabi-Yau appear instead in the twisted superpotential $W$
(see section \ref{sec:twisted} for more details) and correspond  in the GLSM
 to the complexified FI parameters
\beq
\tau_a= {\vartheta_a \over 2\pi} +i\xi_a\,.
\eeq
These can be interpreted as background expectation values for twisted chiral multiplets.

Since the superpotential couplings  \rf{superp} on the (squashed) two-sphere are the same as in flat space
and $\CQ$-exact, standard decoupling arguments \cite{Witten:1993yc} show that the GLSM partition
function on the (squashed) two-sphere does not depend on the superpotential.
The only constraint on the superpotential is that it carries $R$-charge $-2$, so that
it can be coupled supersymmetrically to the (squashed) two-sphere.
The partition function of the GLSM on the (squashed) two-sphere is, however,
a non-trivial function of the complexified FI  parameters, which are coordinates on the
quantum K\"ahler moduli space of the Calabi-Yau.

Compact, non-singular Calabi-Yau manifolds  do not admit continous
isometries.\footnote{Non-compact Calabi-Yau manifolds and singular Calabi-Yau manifolds, however,
admit continous isometries. These isometries can be used to define equivariant Gromov-Witten invariants.
For such manifolds, the corresponding GLSM two-sphere partition function is
a function of the twisted masses, which get identified with the equivariant parameters of the nonlinear
sigma model.}  This is realized in the associated superpotential $\cW$ of the GLSM
breaking all the  symmetries of the GLSM kinetic terms. The absence of flavour symmetries   forbids
the addition of twisted masses $m$  for the chiral multipliets.
Since the $R$-charges $-q$ of the fields are
fixed by demanding that the superpotential $\cW$ has $R$-charge $-q_\CW = -2$ and no flavor symmetries,
we conclude that the partition function on the (squashed) two-sphere for compact Calabi-Yau GLSM's
are non-trivial functions of the complexified K\"ahler moduli, as summarized in the notation $Z(\tau_a,\bar \tau_a)$.
This is precisely the data that enters in the exact K\"ahler potential in the quantum K\"ahler moduli space of a compact
Calabi-Yau. This provides a simple consistency check on the conjecture.

As already mentioned in \cite{Jockers:2012dk},
the K\"ahler potential $\cK$ is known from the work of Cecotti and Vafa \cite{Cecotti:1991me}
to compute the inner product  between certain ground states in the $\cN=(2,2)$ superconformal field theory
to which the GLSM flows in the infrared.   This inner product is a  function of the Calabi-Yau K\"ahler  moduli.
These moduli, in turn,  correspond to   marginal operators in the $\cN=(2,2)$ superconformal field theory.
These operators   are superconformal descendants of    operators in the twisted chiral ring of the conformal field theory.
The K\"ahler potential in  the quantum K\"ahler moduli space  is given  by \cite{Cecotti:1991me}
\begin{align}
  \cK(\tau_a,\bar \tau_a)= - \text{log}\  {}_\text{R}\langle \bar 0 | 0 \rangle_\text{R} \ ,
  \label{inner}
\end{align}
where $| 0 \rangle_\text{R}$ denotes the canonical supersymmetric vacuum in the Ramond sector while
$_{\text R}\langle \bar 0 |$ denotes the conjugate to the canonical vacuum.
We   return to this below and    refer to reference \cite{Cecotti:1991me}  for   more details
regarding the relation between the K\"ahler potential and  this overlap of states.

We present in the rest of this section a physical proof of the above conjecture, completed in the
following two steps. We first compute the exact partition function on the squashed
two-sphere $S_b^2$, denoted by $Z_b$, and prove that $Z_b$ is independent of the
squashing parameter $b$. In particular, it equals the partition function $Z$ on the round two-sphere.  In section \ref{sec:LGG}
we find an alternative proof of the conjecture using Landau-Ginzburg models.

We then   show that the path integral on the squashed two-sphere $S^2_b$
in the  degenerate limit $b\to0$ yields a path integral
representation of the overlap $_\text{R}\big\langle \bar 0 | 0 \big\rangle_\text{R}$ of
the canonical ground states in the Ramond sector
of the $\CN=(2,2)$ superconformal field theory. Combining these two results,  we
arrive at the very relation we were aiming to prove
\begin{align}
  Z(\t_a,\bar\t_a) = Z_b(\t_a,\bar \t_a) \overset{b\to0}{=} \ _R\big\langle \bar 0 | 0 \big \rangle_\text{R}\,
  = e^{-\cK(\t_a,\bar \t_a)}\ .
\end{align}

It is well known, however, that the K\"ahler potential $\cK$ is only defined up to K\"ahler transformations.
These act as follows
\beq
\cK(\tau,\bar\tau)\rightarrow \cK(\tau,\bar\tau)+ f(\tau)+\bar f(\bar\tau)\,.
\eeq
This raises a question: what is the gauge theory counterpart of this ambiguity?
This ambiguity is mirrored in the choice of supersymmetric couplings on the (squashed) two-sphere
for the background twisted chiral multiplet fields corresponding to the complexified FI parameters $\tau_a$.
As shown in section \ref{sec:twisted},
the inclusion of a twisted superpotential $W$ for the background twisted chiral multiplet changes the partition
on the (squashed) two-sphere to\footnote{Another minor source of ambiguity arises from the possible mixing
between the gauge charge and the $U(1)$ $R$-charge.}
\beq
Z(\t_a,\bar\t_a)\rightarrow Z(\t_a,\bar\t_a)\exp\left(-4\pi i l W(\tau_a)- 4\pi i l \overline{W}(\bar \tau_a)\right)\,.
\eeq
Therefore, both the K\"ahler potential $\cK$ and the GLSM partition function $Z$ are
only defined up to K\"ahler transformations.

The next two subsections are devoted to proving the two main claims summarized above.

%%%%%%%%%%%%%%%%%%%%%%%%%%%%%%%%%%%%%%%%%%%%%%%%%%%%%%%%%%%%%%%%%%

\subsection{Exact Partition Function}
\label{sec:partition1}

We compute in this section the partition function $Z_b$ of $\cN=(2,2)$ gauge theories  on
the squashed two-sphere $S^2_b$ by supersymmetric localization. We perform the computation for an arbitrary
gauge theory, including GLSM's for arbitrary K\"ahler manifolds. After a   technical
computation, we prove that the partition function $Z_b$  is independent of the squashing parameter.
This result is key in establishing the equivalence of the partition function of $\cN=(2,2)$ gauge theories
on the round $S^2$ with the exact K\"ahler potential in the quantum corrected K\"ahler moduli space of
Calabi-Yau manifolds. To keep the flow of ideas clear, we relegate many of the technical details of the
computation to the Appendix.

In a localization computation, the path integral restricts to the space of  supersymmetric field
configurations annihilated by a choice of supercharge.
As mentioned earlier, the  supercharge $\CQ$ that generates the $SU(1|1)$ supersymmetry algebra is precisely
the same one that was used in \cite{Doroud:2012xw,Benini:2012ui} to localize the partition function $Z$ of
$\cN=(2,2)$ gauge theories on the round two-sphere $S^2$.
Depending on the choice of deformation term (or contour), as shown in \cite{Doroud:2012xw,Benini:2012ui},
the partition function on $S^2$ can be localized to supersymmetric configurations either in the Coulomb branch
or in the Higgs branch. The Coulomb branch configurations appear in the Coulomb phase of the gauge theory,
while the Higgs branch configurations, which correspond to vortices at the north pole and anti-vortices at
the south pole, appear in the Higgs phase of the theory.
This yields two alternative but equal representations of the partition function $Z$:
the Coulomb and Higgs branch representation.

In order to prove that the partition function $Z_b$ on $S^2_b$ is independent of the squashing parameter
$b$ it suffices to localize the partition function on the squashed two-sphere to the Coulomb branch.
Generalizing the analysis of saddle points of \cite{Doroud:2012xw,Benini:2012ui} to the squashed two-sphere,
we find that the path integral localizes onto the moduli space of solutions of the set of equations
\begin{align}
  D_{\hat i} \s_1 = D_{\hat i} \s_2 = 0 \ , \qquad \left[ \s_1 , \s_2 \right] = 0 \,,
  \qquad
  F_{\hat 1\hat 2} + \frac{\s_1}{f(\th)} =0\,.
  \label{saddle}
\end{align}
Nontrivial gauge field configurations can be characterized by their flux through the squashed two-sphere
\begin{align}
  \frac{1}{2\pi} \int_{S^2_b} F = B\,,
\end{align}
where $B$ takes   values in the Cartan subalgebra $\ft$ of $G$ and is GNO-quantized.
The most general smooth solution  to the saddle point equations \rf{saddle} is\footnote{The choice $\k=1$ ($\k=-1$)
corresponds to  the patch covering the entire two-sphere with the south (north) pole omitted.}
\begin{align}
  A = \frac{B}{2} \left(\k - \cos\th \right) d\varphi\ , \qquad
  \s_1 = - \frac{B}{2l}\ , \qquad \s_2 = \s\ ,
  \label{saddlea}
\end{align}
where $\s$ is valued in the Lie algebra of $G$ and subject to the constraint
\begin{align}
  [\sigma, B]=0\,.
\end{align}

It is noteworthy that the vector multiplet Lagrangian (\ref{Lagvec}) and  the chiral multiplet
Lagrangian (\ref{Lagmatter})  are both $\cQ$-exact. More precisely\footnote{We have normalized the Killing
spinors by $\bar\e \gamma^3 \e=1$.}
\begin{align}
  \CL_\text{v.m.} & = - \frac{1}{g^2}
  \d_\cQ \d_{\bar \e} \ \text{Tr}\left( \frac12 \bar \l \g^3 \l - 2 i {\rm D} \s_2 +   {i\over f(\theta)} \s_2^2 \right)\ ,
  \nonumber \\
  \CL_\text{c.m.} & = - \d_\cQ \d_{\bar \e} \left( \bar \psi \g^3 \psi - 2 \bar \phi \s_2 \phi
  - i \frac{q-1}{f(\theta)}\bar \phi \phi \right)\ .
\end{align}
This implies that the partition function of $\cN=(2,2)$ gauge theories on the squashed two-sphere $S^2_b$ is
independent of the super-renormalizable gauge couplings $g$.
Furthermore, the superpotential couplings  \rf{superp}, which must have
$R$-charge  $-2$ to be supersymmetric on $S^2_b$, are also $\cQ$-exact.
Therefore the partition function $Z_b$ does not depend on the complex parameters that
enter in the superpotential. It does depend, however,  on the complexified FI parameters $\tau_a$, and the $R$-charges $-q$ and
twisted masses $m$ associated to the flavour symmetry group $G_F$ of the gauge theory, which capture the isometries
of the associated K\"ahler manifold.

We can compute the partition function by evaluating the Gaussian integral around
the saddle point field configurations (\ref{saddlea})  using as the $\cQ$-exact deformation term
the vector and chiral multiplet Lagrangians $\CL_\text{v.m.}$ and $\CL_\text{c.m.}$, as well as
a $\cQ$-exact regulator term \eqref{theregulatorterm} given in the Appendix.
%as the s  from which we can compute the partition
%function by evaluating the Gaussian integral around the saddle point field configurations (\ref{saddlea}).
The one-loop determinant for the vector multiplet fields
once we take into account the Jacobian factor reducing the integral over $\sigma$ with the constraint
$[\sigma, B]=0$ to an integral over the Cartan subalgebra is
\begin{equation}
  Z^{\text{v.m.}}_{\text{one-loop}}(\s,B) \cdot J(\s,B)
  =\frac{1}{|\cW(H_B)|}\prod_{\substack{\alpha\in\Delta^+}} \left[\left(\frac{\alpha\cdot B}{2l}\right)^{2}
  +\left(\alpha\cdot \s\right)^2 \right]\,.
\end{equation}
$\alpha$ are the roots of the Lie algebra of $G$ and $\cW(H_B)$ is the Weyl group of the symmetry group
$H_B\subset G$ left unbroken by the flux. The  one-loop determinant for a chiral multiplet  transforming in a
representation $\bf R$ of $G$ is
\begin{align}
  Z_{\text{one-loop}}^{\text{c.m.}}(\s,B,m,q)
  & = \prod_{w\in\bf R} (-i)^{w\cdot B} (-1)^{|w\cdot B|/2}  \frac {\Gamma\left({q\over 2} -
  il({w}\cdot \s +m) + {| {w}\cdot B|\over 2}\right)} {\Gamma\left(1 - {q\over 2} + il({w}\cdot \s+m) +
  {| {w}\cdot B|\over 2}\right)}\,,
  \label{chiralloop}
\end{align}
where $w$ are the weights of the  representation $\bf R$ of $G$.

Combining the one-loop determinants  with the classical contribution and  summing over magnetic fluxes $B$ and
integrating over the moduli space parametrized by $\s$, we get that the $S^2_b$ partition function of an $\cN=(2,2)$
gauge theory is given by\footnote{Here $\xi_{\rm ren}$ is the renormalized FI parameter evaluated at $1/r$. See \cite{Doroud:2012xw}.}
\begin{equation}
  \begin{aligned}
    Z_b( \tau_a,\bar \tau_a,M) = &
    \sum_{B}\frac{1}{|\cW(H_B)|} \int_\ft d\s\,  e^{-4\pi i \xi_{\rm ren} l \text{Tr} \s+i\vartheta \text{Tr} B}
    \prod_{\alpha\in\Delta^+} \left[\left(\frac{\alpha\cdot B}{2l}\right)^{2} +\left(\alpha\cdot \s\right)^2 \right]
    \\ & \hspace*{0.3cm}
    \prod_{w\in\bf R} (-i)^{w\cdot B} (-1)^{|w\cdot B|/2}
    \frac{\Gamma\left({q\over 2} - il({w}\cdot \s +m) + {| {w}\cdot B|\over 2}\right)}
    {\Gamma\left(1 - {q\over 2} + il({w}\cdot \s+m) + {| {w}\cdot B|\over 2}\right)}\  .
  \end{aligned}
  \label{semicolumb}
\end{equation}
The partition function depends on the complexified FI parameters $\tau$ and on the mass $m$ and $R$-charges $-q$
of the chiral multiples through the holomorphic combination $M$ \rf{holol}.

The conclusion of this computation is that the partition function on the squashed two-sphere $S^2_b$ is independent of the squashing parameter $b$.
This is  the first result needed in our proof of the main conjecture \rf{kahler}. In particular, the partition function of the round two-sphere
$Z:=Z_{b=1}$ is identical to that of the squashed two-sphere
\beq
 Z( \tau_a,\bar \tau_a,M) = Z_b( \tau_a,\bar \tau_a,M)\,.
\eeq

Next, we use the freedom to change the squashing parameter $b$ to show that the partition function on the squashed two-sphere $S^2_b$
in the limit $b\rightarrow 0$ provides a path integral representation
of the sought after overlap of states $_\text{R}\langle \bar 0 | 0 \rangle_\text{R}$.

%%%%%%%%%%%%%%%%%%%%%%%%%%%%%%%%%%%%%%%%%%%%%%%%%%%%%%%%%%%%%%%%%%

\subsection{Degenerate Limit and Ground State Inner Product}
\label{sec:partition2}

We now show that the partition function on the squashed two-sphere $S^2_b$ in the degenerate limit $b\to 0$
provides the path integral representation of $_\text{R}\langle \bar 0 | 0 \rangle_\text{R}$\,:
\begin{align}
  Z_{b}(\t_a,\bar \t_a) \overset{b\to0}{=}\    _\text{R}\langle\bar 0 | 0 \rangle_\text{R} \ .
\end{align}
Here we focus on GLSM's that flow in the infrared to $\cN=(2,2)$ superconformal field theories, which   describe
nonlinear sigma models on Calabi-Yau manifolds.

Let us first consider the path integral of a two dimensional  $\CN=(2,2)$ theory on a hemisphere.
This path integral produces the  wavefunction for a state
which is prepared at the boundary of the hemisphere. This  state   is in the same cohomology class as the   ground state.
The projection  to the actual ground state can be obtained through the following canonical construction  \cite{Cecotti:1991me}.
We attach to the hemisphere a   long ``neck" to project into the actual ground state.
More precisely, let us imagine that a semi-infinite cylinder is attached to the boundary of the hemisphere.
Introducing such a semi-infinite cylinder evolves the state $|\Psi \rangle $ on the boundary of the hemisphere
 into a state $e^{-HT} |\Psi \rangle$ where $T \to \infty$.
This evolution can be understood as a projection of the state $|\Psi \rangle$ down to the ground state.
Thus the above path integral chooses a distinguished ground state in the Hilbert space,
which is known as the canonical ground state denoted by $|0\rangle_\text{NS}$.
Due to the spin structure on the hemisphere, the state
thus prepared is in the anti-periodic or Neveu-Schwarz sector.

In order to obtain a state the Ramond   sector rather than the Neveu-Schwarz sector,
we need to  perform   spectral flow on the Hilbert space based on the boundary of the hemisphere.
Cecotti and Vafa \cite{Cecotti:1991me}   introduced an elegant way to implement  spectral flow
by topologically twisting the theory on the hemisphere. That is, they   considered the topological version of
the  same path integral. This can be implemented by  introducing a background connection  $V$ coupled to
the $U(1)$ vector R-symmetry current of the theory. This background gauge field is set to be one-half of
the spin connection $V=\frac12  \w$, thus producing the $A$-twisted theory.
The background connection introduces a non-trivial holonomy when a fermion is transported
around the boundary of the hemisphere. The acquired holonomy is
\begin{align}
  e^{ i\int_{S^1} V }  =
  e^{ i\int_\text{hemisphere} dV } = -1\ ,
\end{align}
which accounts for the very spectral flow mapping states from the Neveu-Schwarz sector
to those in the Ramond sector. As a consequence, the canonical vacuum state $|0\rangle_\text{NS}$
flows to a unique supersymmetric ground state in the Ramond sector denoted by $|0\rangle_\text{R}$.
A very similar construction exists for the conjugate state $_\text{R}\bra {\bar 0}$.  In this case we need to consider the
$\bar A$-twisted theory, obtained by turning on the background expectation value $V=-\frac12  \w$.
Therefore, the path integral representation of the overlap $_\text{R}\langle \bar 0 | 0 \rangle_\text{R}$ corresponds
to joining these two hemispheres through a very long cylinder. The ``fusion" of
these states  and the    tt$^*$ equations \cite{Cecotti:1991me}  were used to
derive  the identity \rf{inner}. We refer to \cite{Cecotti:1991me} for   the details of the derivation.

 We now  argue that   the path integral of $\CN=(2,2)$ supersymmetric theories
on the squashed two-sphere $S_b^2$ in the degenerate limit $b\to 0$
introduces an alternative realization of the spectral flow changing
the ground state in the Neveu-Schwarz sector to the canonical ground state in the Ramond sector. Moreover, it gives   an explicit
path integral representation of the sought after overlap of ground states $_\text{R}\langle \bar 0 | 0\rangle_\text{R}$.

\begin{figure}[t]
  \begin{center}
  \includegraphics[scale=1.7]{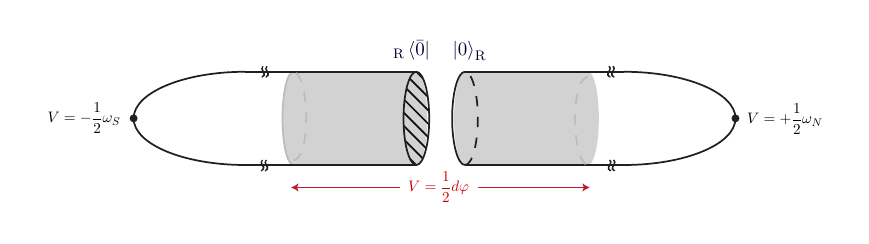}
  \end{center}
  \caption{In the degenerate limit $b\to 0$, the two-sphere is deformed into a union
  of two infinitely stretched cigar-like geometries.
  The shaded region represents an infinitely long flat cylinder on which the standard $\cN=(2,2)$ supersymmetric theories
  in the Ramond sector are defined.}
  \label{cigar}
\end{figure}
In the limit $b\to0$, the sphere gets deformed into a long cylinder with caps at both ends.
It can be described as the  union of two infinitely stretched cigar geometries attached to each other, as
depicted in Fig. \ref{cigar}. In particular, one has a cylinder region arising from zooming near the equator of
the squashed two-sphere $S^2_b$.
In order to make this very explicit, it is instructive to change coordinates to $\rho=\tilde l \cos \theta$,
where  the metric of $S^2_b$  becomes
\begin{align}
  ds^2=d\rho^2+\left( {l\over \tilde l}\right)^2 {\big(  {\rho/ \tilde l} \big)^2\over
  1-\big(  {\rho/\tilde l} \big)^2} d\rho^2+ l^2\left( 1-\big(  {\rho/\tilde l} \big)^2\right) d\varphi^2\,.
\end{align}
The cylinder metric
\beq
ds^2\simeq d\rho^2+l^2d\varphi^2
\eeq
indeed appears in the region $ {\rho/ \tilde l} \ll1$ in the limit $b\to0$,
close to the equator of $S^2_b$.
In the cylinder region the spin connection is trivial $\w \simeq 0$, as expected.

In the flat cylinder region, however, the background connection $V$ for the $U(1)$ $R$-symmetry \rf{backgrounda}
becomes in the limit $b \to 0$
\beq
  V \simeq \frac 12 d\varphi\,.
  \label{backcyl}
\eeq
This has important consequences for the boundary conditions of the spinors around the cylinder.

Recall that an antiperiodic spinor $\e (\varphi)$ on a cylinder with coordinates $(\rho,\varphi)$
in the presence of a background flat connection $V=\nu d\varphi$ is equivalent to a spinor with periodicity
$\e(\varphi+2\pi)=- e^{2\pi i \nu} \e (\varphi)$ without a background gauge field. Therefore, the background gauge field \rf{backcyl}
implies that the boundary conditions of the Killing spinors \rf{killspinQ} on $S^2_b$      around the  cylinder are periodic.
Stated differently, the holonomy induced by the background connection $V$ provides the spectral flow changing
the states based on the boundary of the cigar from the Neveu-Schwarz sector to  the Ramond sector.
Consequently the $\cN=(2,2)$ gauge theory in the flat cylinder region is in the Ramond sector and on the boundary one has
the canonical vacuum $|0\rangle_\text{R}$ after the evolution along the infinitely long cylinder.

In short, the background gauge field $V$ \rf{backgrounda} on the squashed two-sphere allows us to deform the theory on
the hemisphere to the theory on the infinitely long cigar geometry in a supersymmetric fashion.
The infinite squashing limit  $b\to0$ projects the state based on the boundary of the hemisphere down
to the supersymmetric ground state
%$|0\rangle_{\text NS}$,
and simultaneously implements the spectral flow to yield the canonical ground state $|0\rangle_\text{R}$ in the Ramond sector. Therefore, the
path integral on the $S_b^2$ in the limit $b\to 0$ -- which we denoted by  $Z_b$ --  becomes the path integral
representation of the overlap $_\text{R}\langle \bar 0 | 0 \rangle_\text{R}$.

It is noteworthy that, very near the poles, the background gauge field $V$  becomes that required to perform the $A$-twist
and $\bar A$ twist near the north and south poles respectively. In order to analyze these regions $\th\simeq 0$ or $\th\simeq\pi$, we should
first choose vielbein that are regular in a patch around each pole. The corresponding spin connection is then given by
\begin{align}
  \w_{N} = + \left( 1 - \frac{l \cos\th}{f(\th)} \right) d\varphi\ ,
  \qquad
  \w_{S} = - \left( 1 + \frac{l \cos\th}{f(\th)} \right) d\varphi\ .
\end{align}
In the infinitely squashed limit $b\to 0$, one can show that
\begin{align}
  V_N \simeq \frac12 \w_N \ , \qquad V_S \simeq -\frac12 \w_S\ ,
\end{align}
which accounts for the A-twisting near the north pole and $\bar{\rm A}$-twisting
near the south pole. One can thus say that the theory on the squashed two-sphere
nicely interpolates in a supersymmetric manner the A-twisted theory very near the poles with  the $\CN=(2,2)$
theory in the R sector on the flat cylinder.

 We   present a different  proof of the conjecture \rf{kahler}   in the next section in the context of  $\cN=(2,2)$
Landau-Ginzburg models on the two-sphere,  to  which we now turn.

%%%%%%%%%%%%%%%%%%%%%%%%%%%%%%%%%%%%%%%%%%%%%%%%%%%%%%%%%%%%%%%%%%
%%%%%%%%%%%%%%%%%%%%%%%%%%%%%%%%%%%%%%%%%%%%%%%%%%%%%%%%%%%%%%%%%%

\section{Mirror Symmetry  and  the $S^2$ Partition Function}
\label{sec:mirror}

A well established  method for computing worldsheet instantons and Gromov-Witten invariants is mirror symmetry.
In this approach, initiated in \cite{Candelas:1990rm},  the non-perturbative corrections in K\"ahler moduli space
are captured by a purely geometrical problem  in the mirror Calabi-Yau manifold.

In mirror symmetry, the K\"ahler moduli  of a Calabi-Yau get exchanged with the complex structure moduli of the mirror Calabi-Yau.
In the language of the GLSM's,  this corresponds to exchanging chiral multiplets and twisted chiral multiplets,
which as described earlier, parametrize the complex structure and K\"ahler   moduli respectively.

While no general methods are available to determine the mirror manifold to a given Calabi-Yau,
powerful techniques have been developed to compute Gromow-Witten invariants for complete intersections in toric varieties.
In the language of GLSM's, this class of Calabi-Yau manifolds corresponds to   gauge theories based on abelian gauge groups.
In fact, for abelian GLSM's describing non-linear sigma models on target spaces  with non-negative curvature,
Hori and Vafa \cite{Hori:2000kt} devised a powerful ``dualization" method for obtaining the  mirror description of
the corresponding sigma models in terms  of   Landau-Ginzburg models.
This includes the mirrors of sigma models on toric varieties and hypersurfaces in toric varieties.

Given that the partition function of $\cN=(2,2)$ gauge theories on $S^2$ can be computed exactly \cite{Doroud:2012xw,Benini:2012ui},
it is interesting to investigate     mirror symmetry   from the point of view of the partition function on the two-sphere.
This is the goal of this section. By pursuing this line of inquiry, we arrive at an independent proof of the
main conjecture \rf{kahler} using Landau-Ginzburg models. Also, pleasingly, we find that the $S^2$ partition function of
the Hori-Vafa Landau-Ginzburg models \cite{Hori:2000kt} exactly reproduce   the partition function of the  corresponding mirror abelian GLSM's.

Finally, we consider non-abelian GLSM's describing   complete intersections in  Grassmanians, where the
dualization methods used  \cite{Hori:2000kt} are not applicable.
Using the exact result for the gauge theory partition function \cite{Doroud:2012xw,Benini:2012ui},
we prove the  conjectured mirror description put forward    by Hori and Vafa \cite{Hori:2000kt}.

Before presenting these results, we must first construct the mirror $\cN=(2,2)$ supersymmetric  Landau-Ginzburg models
on the two-sphere $S^2$.
These theories are written in terms of twisted chiral multiplets since,
as mentioned above, mirror symmetry exchanges chiral multiplets, which appear in the GLSM description,
with twisted chiral multiplets, which appear in the mirror Landau-Ginzburg description.

%%%%%%%%%%%%%%%%%%%%%%%%%%%%%%%%%%%%%%%%%%%%%%%%%%%%%%%%%%%%%%%%%%

\subsection{Supersymmetric Twisted Chiral Multiplets on S$^2$}
\label{sec:twisted}

The field content of a twisted chiral multiplet is the same as that of the chiral multiplet. We denote the fields in the twisted chiral multiplet by
\beq
 \text{twisted chiral multiplet}: (Y,\overline Y, \chi,\bar\chi,G,\bar G)\,.
\eeq

The $SU(2|1)$ invariant Lagrangian of a twisted chiral multiplet on the round two-sphere of radius $r$ is given by
\begin{align}
  \CL_\text{twisted} = & D^i \overline Y D_i Y + i \bar \chi \g^i D_i \chi +
  \left(\bar G +  \frac{\D}{r} \overline Y\right) \left( G + \frac{\D}{r} Y \right)\,.
  \label{twistedkin}
\end{align}
This action  is invariant under the supersymmetry transformations given by
\begin{align}
  \d Y = & + i  \bar \e  \frac{1 - \g^3}{2} \chi - i  \e  \frac{1 + \g^3}{2} \bar \chi   \nonumber
  \\
  \d \overline Y = & - i   \bar \e \frac{1 + \g^3}{2} \chi + i  \e \frac{ 1 -  \g^3}{2} \bar \chi \
  \nonumber \\
  \d \chi = & + \g^i \frac{ 1 + \g^3}{2}  \e D_i Y -  \g^i \frac{1- \g^3}{2} \e D_i \overline Y
  -  \frac{ 1 + \g^3}{2} \e \left( \bar G + \frac{\D}{r} \overline Y \right)
  - \frac{1 - \g^3}{2} \e \left( G + \frac{\D}{r} Y \right) \
  \nonumber \\
  \d \bar \chi = &  +  \g^i \frac{1 + \g^3}{2} \bar \e D_i \overline Y
  - \g^i \frac{1- \g^3}{2} \bar \e D_i Y
  + \frac{ 1 + \g^3}{2} \bar \e \left( G + \frac{\D}{r} Y \right)
  + \frac{1 - \g^3}{2} \bar \e \left( \bar G + \frac{\D}{r} \overline Y \right) \
  \nonumber \\
  \d G = & - i\bar \e \g^i \frac{1 - \g^3}{2} \nabla_i \chi + i \e \g^i \frac{1 + \g^3}{2} \nabla_i \bar \chi
  - i\frac{\D}{r} \bar \e \frac{1 - \g^3}{2} \chi + i \frac{\D}{r} \e \frac{ 1+ \g^3}{2} \bar \chi\
  \nonumber \\
  \d \bar G = & - i \bar \e \g^i \frac{1 + \g^3}{2} \nabla_i \chi + i \e \g^i \frac{1 - \g^3}{2} \nabla_i \bar \chi
  + i\frac{\D}{r} \bar \e \frac{1 + \g^3}{2} \chi - i \frac{\D}{r} \e \frac{1 - \g^3}{2} \bar \chi \,
  \label{tSUSY}
\end{align}
where $\epsilon$ and $\bar\epsilon$ are Killing spinors on $S^2$.
These transformation realize the $SU(2|1)$ algebra off-shell on the twisted chiral multiplet fields.
The  parameter $\D$ can be identified as the Weyl weight of the twisted chiral multiplet.
We note that the above Lagrangian and supersymmetry transformations  are identical to those of the vector multiplet on $S^2$
once various component fields are identified as follows
\begin{gather}
  Y = \s_2 + i\s_1 \ , \qquad \overline Y = \s_2 - i\s_1 \ , \qquad
  \chi = \l \ , \qquad \bar \chi = \bar \l\nonumber \\
  \qquad G =  {\rm D} - \frac{\s_2}{r}  +iF_{12} \ , \qquad
  \bar G =   {\rm D} - \frac{\s_2}{r}  - i F_{12} \ ,
\end{gather}
and the Weyl weight is fixed to the canonical value, i.e., $\D=1$.

Twisted   superpotential  couplings for the twisted  chiral multiplet can be written on $S^2$ in an
$SU(2|1)$ invariant way in terms of a holomorphic function $W(Y)$. The interaction terms
\begin{align}
  \CL_\text{twisted} = - i  W'(Y) \left( G + \frac{\D}{r} Y \right)
  -   W''(Y) \bar \chi\left( \frac{1- \g^3}{2}\right) \chi  +  \frac ir W(Y)\
  \label{tsuperp}
\end{align}
are invariant under the  supersymmetry transformations above \rf{tSUSY}. Unlike the superpotential  $\cW$ couplings  for chiral superfields in \rf{superp},   the twisted superpotential $W$ couplings \rf{tsuperp} are modified   in comparison to  those   in flat space. As we shall show, the deformation term
\beq
 \frac ir W(Y)
\eeq
in  \rf{tsuperp}    controls most of the  features  of the partition function on the two-sphere even though this term disappears from the action  in the flat space, $r\rightarrow \infty$ limit.

Finally, we note that even though twisted chiral multiplets cannot be minimally coupled to conventional vector multiplets, the field strength multiplet $\Sigma$, which is a twisted chiral multiplet, can couple with a twisted chiral multiplet through a twisted superpotential
\beq
W(\Sigma, Y)\,.
\eeq
In the context  mirror symmetry, the coupling of a twisted chiral multiplet $Y$ to $\Sigma$ via  a dynamical FI term
\beq
W(\Sigma, Y)=\Sigma\, Y
\eeq
is known to play an important role \cite{Hori:2000kt}, and will appear later in our analysis.

Everything that we have discussed for the round two-sphere, admits a fairly simple extension to the squashed two-sphere.

%%%%%%%%%%%%%%%%%%%%%%%%%%%%%%%%%%%%%%%%%%%%%%%%%%%%%%%%%%%%%%%%%%
\subsection{Landau-Ginzburg  Proof of   the Conjecture}
\label{sec:LGG}

We first proceed to the   computation of the $S^2$ partition function of Landau-Ginzburg models
by supersymmetric localization. Using the choice of localizing  supercharge $\CQ$ in \rf{choice},
the path integral localizes to  the saddle point  configurations
\begin{align}
  Y = x + i y \ , \qquad G + \frac{\D}{r} Y = 0 \ ,
  \label{tsaddle}
\end{align}
where $x, y$ are real constants over the two-sphere.

The twisted chiral kinetic terms (\ref{twistedkin}) can be shown to be $\CQ$-exact, and thus can be used as
the $\CQ$-exact deformation term to compute the one-loop determinant
around the saddle points \rf{tsaddle}.
The one-loop determinant of  a twisted chiral multiplet is trivial,  in the sense
that it is independent of the  choice of saddle point configuration $x$ and $y$.

Therefore, the $S^2$ partition function $Z$ of an $\cN=(2,2)$ Landau-Ginzburg model with
a twisted superpotential $W$ reduces to the following integral
\begin{align}
  Z= \int dY d\overline Ye^{ - 4\pi i r W(Y) - 4\pi i r \overline{W}(\overline Y)} =
  \int d x\,  d y \  e^{ - 4\pi i r W(x+iy) - 4\pi i r \overline{W}(x-iy) }  \,.
  \label{twistpartition}
\end{align}
The partition function depends only on the choice of twisted superpotential $W$.
%This result reveals a fact that the two-sphere
%partition function of the theories for twisted chiral multiplets is as an
%observable which is insensitive to D-term variations. (keep this statement here for a while?)

This computation offers a new way to derive the conjecture in \rf{kahler}.
We first quickly recall that Landau-Ginzburg models of the
type we have discussed capture  the physics of   non-linear sigma models on Calabi-Yau target space
in some regions in K\"ahler moduli space.
This fact can be most elegantly seen using the GLSM description of the Calabi-Yau moduli space.
In this framework, the Landau-Ginzburg description corresponds to a different ``phase" \cite{Witten:1993yc}
of the phase diagram parametrized by the K\"ahler moduli.
Furthermore,   Landau-Ginzburg models are mirror to GLSM's describing the nonlinear sigma models
on the mirror Calabi-Yau manifold.

Let us now consider a Landau-Ginzburg description of a two dimensional $\cN=(2,2)$ superconformal
field theory that  can appear at a special point in K\"ahler moduli space.
Consider deforming the conformal field theory by exactly marginal operators, which are
descendants of operators in the twisted chiral ring.
This deformation is implemented in the Landau-Ginzburg description by turning on the twisted superpotential
\beq
  W=i\sum_a \tau_a \cO_a(Y)\,,
\eeq
where $ \cO_a(Y)$ are operators in the twisted chiral ring.
The parameters $\tau_a$ are dimensionless and parametrize the moduli space of the conformal field theory.
The partition function on $S^2$ is a function of these parameters.

We now differentiate the $S^2$ partition function and obtain the following correlation function\footnote{Here we assume that the one-point  function of exactly
marginal operators vanishes, after possibly adding   suitable local counterterms. In \cite{Closset:2012vg} the authors have pointed out that   superconformality and unitarity
are not respected in the framework of the exact three-sphere partition function \cite{Kapustin:2009kz,Hama:2010av,Jafferis:2010un}
due to unitarity violating terms inevitable when putting rigid supersymmetric theories on a curved manifold, even though
the theories are expected to flow to CFTs at infrared. Only after
adding suitable contact terms, the superconformality and unitarity are restored, which imply 
the vanishing of the aforementioned one-point functions. Such contact terms restoring   superconformality may be needed    in
the framework of the two-sphere partition function, which guarantees that the one-point functions
vanish and that the two-point functions in \rf{equalityr} are the same as those in the CFT.}
\begin{align}
  \partial_a \partial_{\bar b} \log Z =
  {\left \langle \mathfrak{O}_a  \cdot \bar{\mathfrak{O}}_{\bar b}  \right \rangle_{S^2}\over Z}\,,
  \label{Zmetric}
\end{align}
where explicitly\footnote{Here we set the radius   $r=1$ for simplicity.}
\begin{align}
  \mathfrak{O}_a & = \int d^2x\sqrt{g} \
  \left( - \partial_i \CO_a \Big( G^i + \D Y^i \right) + \CO_a + \text{fermionic terms} \Big)
  \nonumber \\
  \bar{\mathfrak{O}}_{\bar b} & = \int  d^2x\sqrt{g}  \
  \Big( + \partial_{\bar j} \bar{\CO}_{\bar b} \left( \bar{G}^{\bar j} + \D \bar{Y}^{\bar j} \right) -
  \bar{\CO}_{\bar b} + \text{fermionic terms} \Big)\,.
\end{align}
It is important to now note that   the  operators $\mathfrak{O}_a$
and $\bar{\mathfrak{O}}_{\bar b}$ are invariant under the supersymmetry  transformations \rf{tSUSY}.
Therefore, their correlation functions can be computed by the same localization procedure as we did for the
$S^2$ partition function. Evaluating these operators  on the saddle point field configurations \rf{tsaddle},
we find that the two-point correlation function on $S^2$ \rf{Zmetric} can be reduced to the following simple integral
\begin{align}
  \partial_{a} \partial_{\bar b} \log Z  = -{1\over  Z } \int d\vec x\, d\vec y \ \CO_a(\vec x+ i \vec y)
  \bar{\CO}_{\bar b}(\vec x-i\vec y) e^{-4\pi i W - 4\pi i {\overline W}} \ .
  \label{Zmetric2}
\end{align}

Let us now consider the  insertion of a twisted chiral operator $\CO_a$ at the north pole
and a twisted anti-chiral operator $\bar{\CO}_{\bar b}$ at the south pole of the $S^2$.
These operators preserve precisely the supercharge $\CQ$ \rf{choice} with respect to which
we have localized the $S^2$ partition function.
This implies that
we can compute this two-point correlation function exactly using the same
saddle points. We obtain that
\begin{align}
  \left \langle \CO_a(N) \bar{\CO}_{\bar b}(S) \right\rangle_{S^2}
  = \int d\vec x\,  d\vec y \ \CO_a(\vec x+ i \vec y)
  \bar{\CO}_{\bar b}(\vec x-i\vec y) e^{-4\pi i W - 4\pi i{\overline W}}\ .
  \label{twopoint}
\end{align}

Comparing the    results (\ref{Zmetric}) and (\ref{twopoint}) of the two computations
we have just   performed leads to the following relation
\begin{align}
  - \partial_a \partial_{\bar b} \log Z =
  \frac{\left \langle \CO_a(N) \bar{\CO}_{\bar b}(S) \right\rangle_{S^2}}{Z}\ .
  \label{equalityr}
\end{align}
As   shown by Cecotti-Vafa in \cite{Cecotti:1991me}, the two-point correlation function
on S$^2$ in the r.h.s. of \rf{equalityr} turns out to be the normalized ground state metric $G_{a\bar b}$
in the space of supersymmetric ground states in the Ramond sector
\begin{align}
  \frac{\left \langle \CO_a(N) \bar{\CO}_{\bar b}(S) \right\rangle_{S^2}}{Z}
  = \frac{_\text{R}\langle \bar b | a \rangle_\text{R}} {_\text{R}\langle \bar 0 | 0 \rangle_\text{R} }
  \equiv G_{a\bar b}\ .
\end{align}
Furthermore, using the $tt^*$ equations, it can  be proven  that \cite{Cecotti:1991me}
\begin{align}
  G_{a\bar b} = - \partial_a \partial_{\bar b}\ \log\   _\text{R}\langle \bar 0 | 0 \rangle_\text{R}\ .
\end{align}
Therefore, up to an ambiguity due to K\"ahler transformations, we arrive at the desired result
\begin{align}
  Z(\tau_a,\bar \tau_a) =\   _\text{R} \langle \bar 0 | 0 \rangle_\text{R}=e^{-\cK(\tau_a,\bar \tau_a)}\ .
\end{align}
%

%%%%%%%%%%%%%%%%%%%%%%%%%%%%%%%%%%%%%%%%%%%%%%%%%%%%%%%%%%%%%%%%%%
\subsection{Mirror Symmetry and Hori-Vafa Conjecture}

Hori and Vafa  \cite {Hori:2000kt}  put forward   a powerful method for constructing the mirror
description of non-linear sigma models on K\"ahler manifolds admiting an abelian GLSM description.
This method applies to a broad class of K\"ahler manifolds with non-negative curvature.
These authors provide the mirror description  of sigma models on toric varieties and hypersurfaces
in toric varieties.

In this section we show that  the $S^2$ partition function of  proposed  mirror
Landau-Ginzburg models and abelian GLSM's
are   identical in all cases, even though the  expression for their partition functions are
rather distinct (\rf{twistpartition} versus \rf{semicolumb}).
Furthermore, by studying the partition function of non-abelian GLSM's describing complete intersections
in Grassmanians, we prove a conjecture by Hori and Vafa regarding the mirror Landau-Ginzburg description
of this interesting class of nonlinear sigma models.

We now consider  the partition function of mirror Landau-Ginzburg models for the various cases: toric varieties, hyper surfaces in toric varieties and non-abelian GLSM's.

\medskip
\noindent
{\bf Landau-Ginzburg Mirror of Toric Varieties:}
\medskip

For simplicity, we begin with  toric varieties based on a  GLSM with $U(1)$
gauge group and    $n$ chiral multiplets $\Phi_a$ with  charges $Q_a$, where $a=1,2,\ldots,n$.
We also start with the case of a GLSM with  no  superpotential.

According to the Hori-Vafa  prescription \cite{Hori:2000kt}, the dual Landau-Ginzburg description
involves a twisted chiral multiplet $\S$, the field strength multiplet  constructed from the
vector multiplet,  and $n$ neutral twisted chiral multiplets $Y^a$. The  imaginary part  of $Y^a$ is
periodic with period $2\pi$.
The mirror Landau-Ginzburg  model has a twisted superpotential of the Toda type
and   the twisted chiral multiplets $Y^a$ act as dynamical FI parameters.
The explicit twisted superpotential is
\begin{align}
  W = - \frac{1}{4\pi} \left[  \S \left( \sum_{a=1}^n Q_a Y^a + 2\pi i \tau(\m) \right)
  + i \m \sum_{a=1}^n e^{-Y^a} \right]\ ,
  \label{bbb}
\end{align}
%
%careful at the charge signs
where $\mu$ is an infrared scale, set to be $1/r$ on the two-sphere.

If the original twisted chiral multiplets $\Phi^a$ have twisted masses $m_a$ and $U(1)_R$ charges $-q_a$,
the mirror Landau-Ginzburg model acquires an extra linear twisted superpotential coupling
\beq
W_{\text{lin}}= -\frac{1}{4\pi}  \sum_{a=1}^n\left( m_a + i \frac{q_a}{2r} \right) Y^a\,.
\eeq
In order not to clutter formulas we will mostly focus on the case $m_a=0$ and $q_a\to0^+$, 
the later limit arising due to convergence, as    will be discussed shortly.
Adding it is straightforward and in all cases reproduces the the effect of
turning on these parameters on the mirror GLSM.

Using \rf{twistpartition} and the saddle points of the vector multiplet \rf{saddlea},
the two-sphere partition function of  this Landau-Ginzburg model
is given by\footnote{Note that as stated above, introducing twisted masses and $R$-charges
is easily accomplished  by the simple shift $\sigma Q_a\rightarrow \sigma Q_a+m_a+ i \frac{q_a}{2r}$.}
\begin{align}
  Z = \sum_{B\in \mathbb{Z}} \int_{-\infty}^{\infty} d \sigma \prod_{a=1}^n
   \int_{-\infty}^{\infty}  d x^a \int_{-\pi}^{+\pi} d y^a \
  e^{+ 2 i r\s \left( Q_a x^a - 2\pi \xi \right) + i B \left(
  Q_a y^a + \vartheta \right)} \cdot e^{2 i e^{- x^a} \sin y^a}\,.
\end{align}
Note here that, on the saddle point field configurations,
the imaginary part $-\frac{B}{2r}$ of the field strength superfield $\S$ is quantized ($B \in \mathbb{Z}$)  while the imaginary  part $y^a$ of $Y^a$
is periodic with period $2\pi$.
%For convenience, set $r=1$.

Using the integral representation of the Bessel function of the first kind $J_\a$,
\begin{align}
  J_\a(x) = \frac{1}{2\pi} \int^{+\pi}_{- \pi} dy \ e^{-i\a y} e^{ix\sin y}  \ ,
\end{align}
we can rewrite the partition function as
follows\footnote{In this section we ignore overall irrelevant numerical constants.}
\begin{align}
  Z =\sum_{B \in \mathbb{Z}} e^{+ i B \vartheta}\int d \sigma \ e^{-4\pi i r \s \xi}
  \prod_{a=1}^n (-1)^{B Q_a} \int d x^a \ e^{2 i r\s Q_a x^a}
  J_{BQ_a} \left(2 e^{-x^a}\right)\,.
\end{align}
Interestingly, the Fourier transformation of the Bessel function of
the first kind with a fine-tuned parameter can be expressed in terms of a ratio of Euler gamma
functions
\begin{align}
  \int_{-\infty}^{+\infty} dx \ e^{- q x} e^{2 i Q p  x} J_{B Q}(2 e^{-x}) = (-1)^{\frac{|BQ|- BQ}{2}}\frac12
  \frac{\G\left( \frac q2 + \frac12\left|B Q \right| - i Q p \right)}
  {\G\left( 1 - \frac q2 + \frac12\left|B Q\right| + i Q p \right)}\ .
\end{align}
This identity holds for all nonzero integer $BQ$ and arbitrary $q$.
For $B=0$, one has to be careful about the convergence of the integral. The integral becomes convergent for any positive real $q$.
As explained in \cite{Doroud:2012xw}, the physical origin of the divergence at $q=0$ is due to an extra supersymmetric zero mode.
Using this formula, we can express the partition function of the Landau-Ginzburg model  in the following form
\begin{align}
  Z=\sum_{B \in \mathbb{Z}} e^{+ i B \vartheta}\int d \sigma \
  e^{-4\pi i r \s \xi}
  \prod_{a=1}^n (-1)^{\frac{ | B Q_a| + B Q_a}{2}}
  \frac{\G\left( \frac12\left|B Q_a \right| - ir  Q_a \s \right)}
  {\G\left( 1+ \frac12\left|B Q_a \right| + ir Q_a  \s \right)}\,.
\end{align}
This result  is in perfect agreement with the two-sphere partition function of the mirror  $U(1)$
gauge theory with $n$ chiral multiplets of charge $Q_a$ \rf{semicolumb}.

This matching    can be easily generalized to   GLSM's  with   gauge group $G=\prod_i U(1)_j$
coupled to $n$ chiral multiplets $\Phi^a$ carrying charges encoded in the charge matrix $Q^j_a$.
In this case, the twisted   superpotential of the mirror Landau-Ginzburg  description is given by
\begin{align}
  W = - \frac{1}{4\pi} \bigg[ \sum_{j=1}^{\text{rk}(G)} \S_j \left( \sum_{a=1}^n
  Q_a^j Y^a + 2\pi i \t^j(\m) \right) + i \m \sum_{a=1}^n e^{-Y^a} \bigg]\ .
\end{align}

Extending our previous computation, we
find that  the two-sphere partition function of this   Landau-Ginzburg  model
can be recast into the following form
\begin{align}
  Z = \sum_{B_j \in \mathbb{Z}} e^{+ i B_j \vartheta^j} \int \prod_{j=1}^{\text{rk}(G)} d \sigma_j \  e^{-4\pi i \s_j \xi^j}
  \prod_{a=1}^n (-1)^{\frac{\left|B_j Q^j_a\right| + B_j Q^j_a}{2}}
  \frac{\G\left( \frac12\left|B_j Q^j_a \right| - i r Q^j_a \s_j \right)}
  {\G\left( 1+ \frac12\left|B_j Q^j_a \right| + i r Q^j_a \s_j\right)}\,.
\end{align}
This  exactly reproduces   the partition function  \rf{semicolumb} of the mirror GLSM.

\medskip
\noindent
{\bf Landau-Ginzburg Mirror of Hypersurfaces in  Toric Varieties:}
\medskip

Hypersurfaces
in toric varieties   admit a GLSM description as an abelian gauge theory with a superpotential $\CW$
for the chiral multiplets. As argued in \cite{Hori:2000kt}, the inclusion of a superpotential in the GLSM does not modify the
expression for the twisted superpotential in the mirror Landau-Ginzburg model. There is, nevertheless, an important
difference in the Landau-Ginzburg model for the GLSM without and with a superpotential. The choice of fundamental  coordinates
in the Landau-Ginzburg model changes between the two cases.
  More precisely, adding superpotential in the GLSM affects the topology
of field space in the mirror  Landau-Ginzburg  model from product of cylinders to product of complex planes
with a certain orbifold action.
It is therefore interesting to investigate  how the two-sphere partition function implements
the change of topology of field space after introducing a superpotential term.

For clarity, let us consider the two-sphere partition function of a GLSM with a $U(1)$ gauge group and $n$ chiral multiplets $\Phi^a$
with  charges $Q_a$, where $a=1,2,\ldots, n$. Recall that adding a superpotential $\cW$ on the two-sphere in a supersymmetric
way is only possible if its $R$-charge is $-2$, so we assign $U(1)$ $R$-charges $-q_a$  to the chiral multiplets.
The partition function of this theory is \rf{semicolumb}
\begin{align}
  Z=\sum_{B \in \mathbb{Z}} e^{+ i B \vartheta}\int d \sigma \
  e^{-4\pi i r \s \xi}
  \prod_{a=1}^n (-1)^{\frac{ | B Q_a| + B Q_a}{2}}
  \frac{\G\left( \frac{q_a}{2} + \frac12\left|B Q_a \right| - ir  Q_a \s \right)}
  {\G\left( 1 - \frac{q_a}{2} + \frac12\left|B Q_a \right| + ir Q_a  \s \right)}\ .
  \label{aaa}
\end{align}
Noting that
\begin{align}
  \int_{-\infty}^{+\infty} dx \ e^{- q x} e^{2 i Q p  x} J_{B Q}(2 e^{-x}) = (-1)^{\frac{|BQ|- BQ}{2}}\frac12
  \frac{\G\left( \frac q2 + \frac12\left|B Q \right| - i Q p \right)}
  {\G\left( 1 - \frac q2 + \frac12\left|B Q\right| + i Q p \right)}\ ,
\end{align}
we can rewrite \rf{aaa} into the following form
\begin{align}
  Z & = \sum_{B\in \mathbb{Z}} \int d \sigma \prod_{a=1}^n
  \int  d x^a e^{-q_a x^a} \int_{-\pi}^{+\pi} d y^a \
  e^{+ 2 i r\s \left( Q_a x^a - 2\pi \xi \right) + i B \left(
  Q_a y^a + \vartheta \right)} e^{2 i e^{- x^a} \sin y^a}
  \nonumber \\ & =
  \int d\S \, d\bar \S
  \int \left[ \prod_{a=1}^n dY_a d{\overline Y}_a  e^{-\frac{q_a}{2} (Y^a + {\overline Y}^a )}\right]\
  e^{-4\pi i r W - 4\pi i r \overline{W}}\,,
  \label{finnn}
\end{align}
with the twisted superpotential \rf{bbb}
\begin{align}
  W = - \frac{1}{4\pi} \left[  \S \left( \sum_{a=1}^n Q_a Y^a + 2\pi i \tau(\m) \right)
  + i \m \sum_{a=1}^n e^{-Y^a} \right]\,.
  \label{bbbb}
\end{align}
The final expression \rf{finnn} is almost the same as what one obtained  in our  previous examples
with $\cW=0$, except for extra weight factors $e^{-\frac{q_a}{2} (Y^a + {\overline Y}^a )}$ in the measure.
Although the dual description involves
$n$ twisted chiral multiplets $Y^a$ coupled via the same twisted superpotential \rf{bbbb},
 these extra   factors are responsible for the
 subtle difference in choosing what the fundamental variables of the mirror Landau-Ginzburg model are.
While a twisted chiral multiplet $Y^a$ is the   variable dual to a chiral multiplet $\Phi^a$ with vanishing
$U(1)$ $R$-charge, care must be taken to    choose an appropriate   dual variable for a chiral multiplet $\Phi^a$ of $U(1)$
$R$-charge $q_a$. The proper fundamental variable in this case is (no summation over $a$)
\begin{align}
  X_a = e^{-\frac{q_a}{2} Y^a}\ .
  \label{varrr}
\end{align}
As we shall now  demonstrate, the fundamental Landau-Ginzburg variables \rf{varrr}
that naturally arise from the  two-sphere partition function of the mirror GLSM \rf{finnn}
are in perfect agreement with those proposed by Hori and Vafa  \cite{Hori:2000kt}.

For concreteness, let us consider a degree $n$ hypersurface in $\mathbb{CP}^{n-1}$.
For $n=5$, this hypersurface describes the celebrated  quintic Calabi-Yau threefold.
This Calabi-Yau is described \cite{Witten:1993yc} by an abelian GLSM with $n$ chiral multiplets $\Phi^a$
of gauge charge $+1$ and a chiral multiplet $P$ of gauge charge $-n$. The superpotential is
\begin{align}
  \CW = P \, G_n (\Phi_a) \ ,
\end{align}
where $G_n$ denotes a homogeneous function of $\Phi^a$ of degree $n$. The $R$-charge of $\Phi^a$ is $q=2 \fr$
while the $R$-charge of $P$ is determined by $q=2-2 n \fr $.
 The   parameter $\fr$ can be identified as the gauge charge. Due to the convergence of the integral,
this   parameter should be in the range of $(0,1/n)$.

Looking at \rf{finnn}, the two-sphere partition function of this GLSM can be expressed as
\begin{align}
  Z  = \int d\S\, d\bar \S
  \int  \prod_{a=1}^n \left[dY_a d{\overline Y}_a e^{-\fr(Y_a + {\overline Y}_a)} \right]
  \left[dY_P d{\overline Y}_P e^{-(1-n\fr)(Y_P + {\overline Y}_P)}\right] \
  e^{-4\pi i r W - 4\pi i r \overline{W}}
\end{align}
where the twisted superpotential is given by
\begin{align}
  W = - \frac{1}{4\pi} \bigg[  \S \left( \sum_{a=1}^n Y^a - n Y_P + 2\pi i \tau \right)
  + i \frac1r \Big( \sum_{a=1}^n e^{-Y^a} + e^{-Y_P} \Big) \bigg]\ .
\end{align}
Integration first over $\S,\bar \S$ and then over $Y_P, {\overline Y}_P$
leads to further simplification of the two-sphere
partition function
\begin{align}
  Z = \int \prod_a \left[ d {\widetilde X}_a d \overline{{\widetilde X}}_a \right] \ e^{- W_{\text{eff}}+ \overline{W}_{\text{eff}}} \,,
\end{align}
where the effective twisted superpotential is
\begin{align}
W_{\text{eff}} = \sum_a {\widetilde X}_a^n + e^{-2\pi i \t/n}\prod_a {\widetilde X}_a\ .
\label{quinticW}
\end{align}
Here the canonical  variables ${\widetilde X}_a$ are given by
\begin{align}
  {\widetilde X}_a = e^{-\frac1n Y_a} \,.
\end{align}
Since  the imaginary part  of the twisted chiral field $Y_a$ is periodic with period $2\pi$,
we need  to identify $\widetilde X_a \simeq e^{2\pi i /n} \widetilde X_a$. This implies that
the present model is indeed an orbifold of the Landau-Ginzburg with twisted superpotential \rf{quinticW}.
Note that the above choice of fundamental variables is in perfect agreement with the choice
proposed in \cite{Hori:2000kt}. 
%In other words, the ground state $|0\rangle_\text{R}^c$ of the compact theory
%corresponds to the $|\Sigma\rangle_\text{R}^{nc}$ in the non-compact theory. One can verify explicitly from the two-sphere
%partition functions of the corresponding GLSMs that
%
%\begin{align}
  %Z^c (q_a=0^+, q_P = 2^-) \propto \partial_t \partial_{\bar t} Z^{nc}(q_a=0^+, q_P=0^+) \ \to \
 % _\text{R}^c \langle \bar{0} | 0 \rangle_\text{R}^c \propto ~_\text{R}^{nc} \langle \overline{\Sigma} |\Sigma \rangle_\text{R}^{nc}\ .
%\end{align}
%
%}

\medskip
\noindent
{\bf Landau-Ginzburg Mirror of Nonabelian GLSM's:}
\medskip

The dualization methods in  \cite{Hori:2000kt} to derive mirror Landau-Ginzburg models,
while very powerful for abelian gauge theories, can  not extend to non-abelian gauge theories.
On the other hand, the exact computation of the partition function of $\cN=(2,2)$ gauge theories
on the two-sphere \cite{Doroud:2012xw,Benini:2012ui} applies to arbitrary gauge theories,
and offers a new perspective on the problem of finding the mirror Landau-Ginzburg models to non-abelian GLSM's.
These GLSM's describe interesting non-linear sigma models on hypersurfaces over the
Grassmanian or flag varieties in general, for which no universal method of computation of
Gromov-Witten invariants are known.

For concreteness let us consider  a $U(N)$ gauge theory coupled to a chiral multiplet
in a representation $\mathbf{R}$ of the gauge group.
The two-sphere partition function is given by \rf{semicolumb}
\begin{align}
  Z & = \frac{1}{|\CW(G)|} \sum_{B_j} e^{+ i \vartheta\sum_{j} B_j}
  \int \prod_{j=1}^{\text{rk}(G)}  d \sigma_j \
  e^{-4\pi i  \xi \sum_j \sigma_j} \ \prod_{j<k} \left|\s_j -\s_k - i \frac{ B_j-B_k}{2r}  \right|^2
  \nonumber \\ &  \ \ \  \times
  \prod_{\r^j} (-1)^{\frac{\left|\r^j B_j \right| + \r^j B_j }{2}}
  \frac{\G\left( \frac12\left|\r^j B_j \right| - i r \r^j \s_j \right)}
  {\G\left( 1+ \frac12 \left|\r^j B _j \right| + i r \r^j \s_j\right)}\,.
\end{align}
Here $\r^j$ denote the weights of the representation $\bf R$ of $U(N)$.
Using the various identities presented above, we can rewrite the two-sphere
partition function of this non-abelian gauge theory
in the following form
\begin{align}
  Z & =
  \frac{1}{|\CW(G)|} \sum_{B_j} \int \prod_{j=1}^{\text{rk}(G)}  d \sigma_j
  \prod_{\r \in \mathbf{R}}
  \int d x^\r \int_{-\pi}^{+\pi} d y^\r \
  \prod_{j<k} \left|\s_j -\s_k - i \frac{ B_j-B_k}{2r}  \right|^2
  \nonumber \\ & \ \ \ \times
  e^{ + 2 i \s_j \left( \r^j x^\r - 2\pi \xi \right) + i B_j \left(
  \r^j y^\r + \vartheta \right)} \cdot e^{2 i  e^{- x^\r} \sin y^\r}\,.
  \nonumber
  \end{align}
This can be further massaged into the following conducive Landau-Ginzburg form
\beq
  Z=\frac{1}{|\CW(G)|} \int \Big[\prod_{j=1}^{\text{rk}(G)} d\S_j d\bar \S_j \Big]
  \Big[ \prod_{\r \in \mathbf{R}} dY^\r d{\overline Y}^\r\Big]  \prod_{j<k} \left| \S_j - \S_k \right|^2 \
  e^{-4\pi i W - 4\pi i \overline{W}}\,,
\eeq
where the effective twisted superpotential is given by
\begin{align}
  W = - \frac{1}{4\pi} \left[ \sum_{j=1}^{\text{rk}(G)} \S_j \left( \sum_{\r \in \mathbf{R}}
   \r^jY^\r + 2\pi i \t \right) + i \m \sum_{\r} e^{-Y^\r} \right]\ .
\end{align}
This result  agrees perfectly with the conjectured dual description of this  nonabelian GLSM proposed by
Hori and Vafa in \cite{Hori:2000kt} and follows rather directly from the exact results of the partition function of $\cN=(2,2)$ gauge theories on the two-sphere.

%%%%%%%%%%%%%%%%%%%%%%%%%%%%%%%%%%%%%%%%%%%%%%%%%%%%%%%%%%%%%%%%%%
%%%%%%%%%%%%%%%%%%%%%%%%%%%%%%%%%%%%%%%%%%%%%%%%%%%%%%%%%%%%%%%%%%

\section{Discussion}
\label{sec:conclusion}

In this paper we have found  two physical proofs of the conjecture originally put forward  in
 \cite{Jockers:2012dk}. This  conjecture relates the two-sphere partition function
of $\CN=(2,2)$ supersymmetric theories with
the K\"ahler potential $\cK$ on the quantum K\"ahler moduli space of   Calabi-Yau manifolds
\begin{align}
  Z(\tau_a,\bar \tau_a)=e^{-\cK(\tau_a,\bar \tau_a)}\ .
\end{align}
One proof uses the invariance of the two-sphere partition function under squashing, which we establish by computing   the squashed two-sphere partition function. We then note   that this
path integral in the infinitely squashed limit represents
 the overlap  of the canonical ground state  $_\text{R}\langle \bar 0 | 0 \rangle_\text{R}$
of the  infrared  $\CN=(2,2)$ superconformal theory in the Ramond sector  defined on a flat cylinder, which indeed computes the K\"ahler potential $\cK(\t_a, \bar \t_a)$. Here the background gauge field needed to define the theory on the squashed two-sphere
plays a key role in implementing the spectral flow  to the Ramond sector.

  Therefore,  the partition function on $S^2$
 of Calabi-Yau GLSM's  offers a new explicit   method for  computing the K\"ahler potential and Gromov-Witten invariants in Calabi-Yau sigma models. One advantage of this approach is that  it does not discriminate between Calabi-Yau manifolds obtained from
 abelian and non-abelian Calabi-Yau GLSM's.  This is in stark contrast with other approaches, where    powerful, general  methods exist for complete intersections in toric varieties
 -- which correspond to abelian gauge theories -- but no general methods  to compute Gromov-Witten invariants are available otherwise, for example, for Calabi-Yau manifolds based on non-abelian gauge theories. Certain non-abelian GLSM's for non-complete intersections have been constructed recently
 \cite{Hori:2006dk,Donagi:2007hi,Hori:arXiv1104.2853,Jockers:2012zr}, and the two-sphere partition function is a new exact method to study them.  Indeed, the virtues of the two-sphere partition function  approach to Gromov-Witten invariants have already been exposed in \cite{Jockers:2012dk}, where  Gromov-Witten invariants for Calabi-Yau manifolds for which no other method is available were computed.

Two dimensional  $\cN=(2,2)$ theories  on the two-sphere can be enriched by adding supersymmetric defects. Among these, line operators  localized at the equator appear  particularly promising. The simplest such defect is a Wilson loop operator, which was already considered in \cite{Doroud:2012xw}.
It would be interesting to construct domain walls   preserving the   $SU(1|1)$ symmetry used to localize the $S^2$ partition function, and to compute the partition function in the presence of the domain wall, similarly to what was done for domain walls \cite{Drukker:2010jp} (see also \cite{Hosomichi:2010vh}) on the four-sphere \cite{Pestun:2007rz}.

Just as the two-sphere partition function computes   worldsheet instantons in a Calabi-Yau, we  expect   that the $S^2$ partition  function in the presence of a domain wall  to compute   woldsheet instanton  in the presence of   D-branes  in the Calabi-Yau geometry.
This is a particularly promising speculation since it is notoriously difficult to compute such worldsheet instantons by conventional methods, while the result of the computation of the partition function in the  presence of a domain wall will be  rather explicit. This may provide a new effective tool to compute open Gromov-Witten invariants in Calabi-Yau geometries.

We have also defined    supersymmetric theories for twisted chiral multiplets on the two-sphere,
and computed  the partition function exactly using   localization techniques.
Our results have found several interesting applications.
Firstly, they provide an alternative proof of
the above conjecture in the context of   Landau-Ginzburg models. Secondly these results can be used
to provide precision studies of mirror symmetry. In particular, our analysis allowed us  to show that the partition function
of the  Landau-Ginzburg  models
 proposed by Hori and Vafa \cite{Hori:2000kt} are in exact agreement with the partition function of the mirror GLSM's. Interestingly, for the case of hypersurfaces in toric varieties,
the two-sphere partition function of the GLSM naturally knows what appropriate  variables should be used in the mirror Landau-Ginzburg  description.
The computation of the two-sphere partition function for arbitrary gauge theories provides us with a
systematic
approach  to construct the mirror description of a   GLSM with a non-abelian gauge group. This is rather interesting since the usual methods to construct the mirror description break down for non-abelian GLSM's.
As an explicit application, we have used our results to get  nontrivial evidence for the mirror description of a certain non-abelian GLSM
describing   complete intersections in the Grassmannian. The mirror Landau-Ginzburg description we have found proves the conjectured description advanced by Hori and Vafa \cite{Hori:2000kt}. It would be interesting to push these ideas as a new tool towards understanding non-abelian dualities in two dimensional gauge theories.

In the GLSM approach to nonlinear sigma models on K\"ahler manifolds, the isometry group $G_F$ of the target space is realized in the GLSM as the flavour symmetry group. Deforming the GLSM by  twisted masses $m$, which take values  in the Cartan of $G_F$,   corresponds to adding a supersymmetric potential to the nonlinear sigma model that gauges the corresponding isometries. The nonlinear sigma model in this case computes equivariant Gromov-Witten invariants with respect to the Cartan of $G_F$. In this context the twisted masses map  to the equivariant parameters. In these theories, the partition function of the GLSM on the (squashed) two-sphere is a non-trivial function of the twisted masses. It would be interesting to explore in detail   the two-sphere partition function for these theories as a method for  extracting equivariant Gromov-Witten invariants.

In  \cite{Doroud:2012xw}\cite{Benini:2012ui} it was shown that the two-sphere partition function of gauge theories admit both a Coulomb branch and Higgs branch representation. In the first representation the partition function is written as an integral over the Coulomb branch, while in the Higgs branch representation the partition function is expressed as a sum over Higgs vacua of the product of vortex and anti-vortex partition functions.\footnote{For some theories, as shown in \cite{Doroud:2012xw}, the Higgs branch representation is a Toda CFT correlator, and the sum over conformal blocks mimics the sum over Higgs vacua in the gauge theory.} We expect -- in specific examples -- a similar  dual description of the partition function of Landau-Ginzburg models with a  twisted superpotential $W$. We have shown that the partition function of  such theories is given by the following integral
\begin{align}
  Z= \int dY d\overline Ye^{ - 4\pi i r W(Y) - 4\pi i r \overline{W}(\overline Y)}\,.
  \label{twistpartitiona}
\end{align}
In section \ref{sec:mirror} we have demonstrated  in   examples how this integral reproduces the ``Coulomb branch" representation of the mirror gauge theory.  The ``Higgs branch" representation of the partition function \rf{twistpartitiona} as a sum of the product of   a holomorphic  and an
anti-holomorphic function of the parameters in the twisted superpotential, can be obtained by decomposing the integral as a sum over integration cycles, which depend on the critical points of the twisted superpotential $W$.  These methods have been  recently  used by  Witten \cite{Witten:2010cx} and Gaiotto and Witten \cite{Gaiotto:2011nm} in the context of analytic continuation of Chern-Simons theory.

The partition function on the squashed two-sphere can be enriched  both by   introducing twisted mass terms
and by inserting twisted chiral operators at the north pole and twisted antichiral operators at the south pole.
This follows from the fact that twisted mass terms and these operator insertions at the poles respect
the supercharge $\cQ$ with which we localized the path integral. It would be particularly
interesting to explore the relation between this  enriched two-sphere partition function
and the ground state metric in the $tt^*$ formulation for   massive theories \cite{Cecotti:1991me}, with an  emphasis
on possible applications to the wall-crossing phenomena of   BPS   kinks   \cite{Cecotti:1992qh}.

Along a similar  line of inquiry, it would  also be interesting to generalize the  ideas presented above
 to four dimensional $\cN=2$ gauge theories, as these theories  realize  many of the same  phenomena we have
discussed in the context of  two dimensional $\cN=(2,2)$ theories.

%%%%%%%%%%%%%%%%%%%%%%%%%%%%%%%%%%%%%%%%%%%%%%%%%%%%%%%%%%%%%%%%%%

\vskip 2cm
\centerline{\bf\large Acknowledgement}
\vskip 5mm
We would like to thank Stefano Cremonesi, Nima Doroud, Davide Gaiotto and Bruno Le Floch for valuable discussions. We   thank the
 Simons Center for Geometry and Physics for hospitality, where this work was initiated and completed, and where a preliminary version of these results was presented.
S.L. would like to thank Aspen Center for Physics (NSF Grant No. 1066293) for hospitality.
Research at the Perimeter Institute is supported in part by the Government of Canada through NSERC and by the Province of Ontario through MRI.
J.G. also acknowledges further support from an NSERC Discovery Grant and from an ERA grant by the Province of Ontario.
The work of S.L. is supported by the Ernest Rutherford fellowship of the Science \& Technology Facilities Council ST/J003549/1.

\newpage

%%%%%%%%%%%%%%%%%%%%%%%%%%%%%%%%%%%%%%%%%%%%%%%%%%%%%%%%%%%%%%%%%%
%%%%%%%%%%%%%%%%%%%%%%%%%%%%%%%%%%%%%%%%%%%%%%%%%%%%%%%%%%%%%%%%%%

\centerline{\Large \bf Appendix}

\appendix

\section{Details of the Computation}
\label{sec:details}

We present in this section the details of the one-loop determinant computations.

Our choice of supercharge $\CQ$ is exactly the same to what is chosen for the round two-sphere, which
generates the $SU(1|1)$ supersymmetry algebra. The Killing spinors associated to this supercharge $\CQ$
are denoted by $\e_\CQ$ and ${\bar \e}_\CQ$, normalized such that
\begin{align}
  {\bar \e}_\CQ \g^3 \e_\CQ = 1\ .
\end{align}
From now on, we ignore the subscript $\CQ$ unless it causes any confusions.
Denoting the rests of bi-linear forms as
\begin{align}
  w= \bar \e \e  \ , \qquad v^i = \bar \e \g^i \e \ ,
\end{align}
it is easy to show that they satisfy the following relations frequently used in what follows
\begin{align}
  w^2 - v_i v^i =1 \ , \qquad v_i \bar \e \g^i + \bar \e \g^3  = w \bar \e\ , \qquad
  v_i \g^i \e + \g^3 \e  = w \e \ ,
  \label{identity3}
\end{align}
and
\begin{align}
  \g^i D_i w = \frac{1}{f} \g^3\g^i v_i\ ,
  \qquad  D_i v^i = 0 \ , \qquad
  \g^{ij} D_i v_j = \frac{2}{f} w \g^3 \ , \qquad
  v^i \partial_i f = 0 \ .
  \label{identity2}
\end{align}
The following identities is also useful
\begin{align}
  \frac14  \Big( \CR + 2 i G_{ij} \g^{ij} \Big) \e & =  \Big(
  \frac{1}{2f^2} -  \frac{1}{2} \g^i \g^3 \partial_i f^{-1} \Big) \e\ ,
  \nonumber \\
  \frac14 \bar \e \Big( \CR + 2 i G_{ij}\g^{ij} \Big) & =
  \bar \e \Big( \frac{1}{2f^2} - \frac12 \g^i \g^3 \partial_i f^{-1} \Big) \ ,
  \label{identity1}
\end{align}
where $G=dV$.

It would be   tedious to work out all the eigenmodes.
Limited to the one-loop determinant, it is in fact not necessary to know the explicit
expressions of all the eigenmodes and their eigenvalues.
This is due to  the vast cancellation between the contributions to the one-loop determinant from
bosonic and fermionic modes   thanks  to supersymmetry.
It is therefore sufficient to understand how this cancellation happens.

\subsection{Matter Multiplet}

First, we consider the one-loop determinant from a chiral multiplet of
R-charge $-q$ and gauge charge one, coupled to an abelian vector multiplet.\footnote{The non-abelian case follows trivially since $[\sigma,B]=0$.}
It turns out that,
instead of the original Lagrangian (\ref{Lagmatter}),
a different choice of a $\CQ$-exact deformation simplifies the one-loop determinant computation.
We choose the regulator Lagrangian as follows
\begin{align}\label{theregulatorterm}
  \CL_\text{reg} = - \d_\e \d_{\bar \e} \left( \bar \psi \g^3 \psi - 2 \bar \phi \s_2 \phi \right) \ ,
\end{align}
which leads to the kinetic operators $\D_b$ and $\D_f$
\begin{align}
  \D_b & = - D_i^2 + \s_1^2 + \s^2  + \frac{q}{4} \CR
  + \frac{q-1}{f} v^i D_i  + \frac{q-1}{f} w \s_1   + \frac{q^2 - 2q}{4f^2}\ ,
  \nonumber \\
  \D_f & =   - i \g^i D_i + i \s_1 - \s \g^3
  - i \frac{1}{2f}\g^3 + i \frac{q-1}{2f}v_i \g^i + i\frac{q-1}{2f} w\ ,
\end{align}
acting on scalars and fermions of R-charges $-q$ and $1-q$, respectively.
Here $\s_1$ and the background gauge field $A$ involved in the covariant derivative
are given by (\ref{saddle}). For simplicity,
it is also useful to consider spinor eigenmodes for an operator defined by
\begin{align}
  \g^3 \D_f \Psi = M \Psi\ ,
\end{align}
rather than those for the original one.

\paragraph{super multiplet}

Given such a spinor eigenmode $\Psi$ for $\g^3 \D_f \doteq M$, %that is,
%%
%\begin{align}
%  \D_f \Psi = M \g^3 \Psi\ ,
%\end{align}
%%
it is straightforward to show that
\begin{align}
  \Phi = \bar \e \Psi\
  \label{map0}
\end{align}
is a scalar eigenmode for $\D_b\doteq - M ( M +2 \s )$.
Secondly, let $\Phi$ denote a scalar eigenmode for
$\D_b\doteq - M( M +2\s )$. Defining a pair of spinors as
\begin{align}
  \Psi_1 = \g^3 \e \Phi \ , \qquad
  \Psi_2 = i \g^i \e D_i \Phi + i \e \s_1 \Phi + \g^3 \e
  \left( \s \Phi+ i \frac{q}{2f} \right)\Phi \ ,
  \label{map1}
\end{align}
one can show that $\gamma^{3}\Delta_{f}$ can act on them as follows
\begin{align}
  \g^3 \D_f \begin{pmatrix} \Psi_1 \\ \Psi_2 \end{pmatrix} =
  \begin{pmatrix} -2 \s & 1 \\ M ( M +2\s) & 0 \end{pmatrix}
  \begin{pmatrix} \Psi_1 \\ \Psi_2 \end{pmatrix} \ .
\end{align}
The matrix on the right hind side has eigenvalues
\begin{align}
  \g^3 \D_f \doteq M \ , \ \ - (M +2\s )\ .
\end{align}
We thus found a pair between a scalar eigenmode with $\D_b \doteq -M ( M +2\s)$
and two spinor eigenmodes with $\g^3\D_f \doteq M, -(M+2\s)$. The map pairing
scalar and spinor eigenmodes could be guessed from the SUSY variation rules (\ref{SUSYmatter}).

Due to the cancellation,  the contribution to the one-loop determinant from any modes participating
in this pairing becomes trivial. There are two types of eigenmodes
which provide nontrivial contribution to the one-loop determinant,
which fail to fall into the aforementioned multiplet structure.
That is to say, either the map (\ref{map0}) or the map (\ref{map1}) becomes ill-defined.

\paragraph{unpaired spinor eigenmodes} We start with the unpaired spinor egenmodes
which vanish when contracted with $\bar \e$. As consequence, these modes do not
have scalar partners. They take the following form
\begin{align}
  \Psi = \bar\e g(\th,\varphi) \ ,
   \end{align}
where $g(\th,\varphi)$ is a scalar of R-charge $2-q$ and of gauge charge $+1$.
Then, the eigenmode equation $\D_f \Psi = \g^3 M \Psi$ gives us
\begin{align}
  - i \g^i \bar \e D_i g + i \s_1 \bar \e g  =
%  & = \left(
%  - \frac{i}{f}  + M  + \s_2 + i\frac{1}{2f} \right) \g^3 \bar \e g
%  - i \frac{q-1}{2f} ( v_m \g^m + w ) \bar \e  g
%  \nonumber \\ & =
  \left( M + \s_2 + i \frac{q-2}{2f} \right) \g^3 \bar \e g
\end{align}
with the covariant derivative defined by
\begin{align}
  D g = d - i \frac{B}{2} \left( \k- \cos\th \right) d\varphi + i\frac{q-2}{2} \left( 1 -
  \frac{l}{f} \right) d\varphi\ , \qquad \s_1 = - \frac{B}{2l}\ .
\end{align}
%
%Componentwise, one can obtain
%%
%\begin{align}
%  -i \cos\frac{\th}{2} \left( \frac{1}{f} D_\th  - \frac{i}{l \sin\th} D_\varphi \right) g
%  + i\sin{\frac{\th}{2}} \s_1  g &
%  = \sin\frac{\th}{2} \left( M + \s + i \frac{q-2}{2f} \right)  g\ ,
%  \nonumber \\
%  -i \sin\frac{\th}{2} \left( \frac{1}{f} D_\th  + \frac{i}{l \sin\th} D_\varphi \right) g
%  + i\cos{\frac{\th}{2}} \s_1  g &
%  = -\cos\frac{\th}{2} \left( M + \s + i \frac{q-2}{2f} \right)  g\ .
%\end{align}
%%
It leads to the following two equations
\begin{align}
  \partial_\varphi g  = - \left( Ml + \s l + i \frac{q-2}{2}  - i \frac{B\k}{2} \right) g \ ,
  \label{spinor1}
\end{align}
and
\begin{align}
  \frac{-i}{f} \partial_\th g + \frac{\cos\th }{\sin\th} \left( M + \s + i \frac{q-2}{2f} \right) g
  + \frac{i}{\sin\th} \s_1 g = 0 \ .
  \label{spinor2}
\end{align}
Assuming $g(\th,\varphi) = e^{-i J \varphi} h(\th)$, the first equation (\ref{spinor1}) determines the eigenvalue $M$,
\begin{align}
  i J = \left( M l  + \s l  + i \frac{q-2}{2} - i \frac{B\k}{2} \right) \ ,
\end{align}
while the second equation (\ref{spinor2}) determines the function $h(\th)$,
\begin{align}
  \frac{-i}{f} \partial_\th h + \frac{\cos\th }{\sin\th} \left( i \frac{J}{l} - i \frac{q-2}{2l} + i \frac{q-2}{2f} + i \frac{B\k}{2l} \right) h
  - \frac{i}{\sin\th} \frac{B}{2l}  h = 0 \ .
\end{align}
It is not important to know the precise expression of the solutions to the above differential equation.
All we need to care about is when the solution becomes non-normalizable. Looking at the differential equation, the solution can develop singularities at the north pole $\th=0$ and the south pole $\th=\pi$.
One can easily show that the solutions near the pole are approximate to
\begin{align}
  h(\th) \sim \sin^J\th \ ,
\end{align}
For the normalizability, one needs to require $J$ to be non-negative. As a consequence,
the eigenvalues for unpaired spinors eigenmodes are given by
\begin{align}
  M l = i \left( J + 1 + i\s l - \frac q2 + \frac{|B|}{2} \right) \ ,
  \label{result1}
\end{align}
where $J \geq 0$\footnote{Note that there is no eigenmode for
$Ml = i \left( J + 1 + i\s_2 l - \frac q2 - \frac{|B|}{2}\right) $ for
$0 \leq J < |B|$ which are well-defined all over the squashed two-sphere.}.

\paragraph{missing spinor eigenmodes}

Other pieces contributing to the one-loop determinant arise from the missing spinor eigenmodes.
It happens when the map (\ref{map1}) does not provide two independent spinor eigenmodes from a
single scalar eigenmode $\Phi$. In other words, $\Psi_1 = - M \Psi_2$ for a constant $M$. One
can show that $\Psi_1=-M\Psi_2$ is indeed a spinor eigenmode,
\begin{align}
   \g^3 \D_f \Psi_1 = -2\s_2 \Psi_1 + \Psi_2 = - \left( M + 2\s_2 \right) \Psi_1 \ .
\end{align}
Note that any scalar function $\Phi$ of R-charge $-q$ satisfying the relation below
\begin{align}
  \Psi_1 = - M \Psi_2 \ \to \ - i\g^i \e D_i \Phi - i \s_1 \e \Phi = \g^3 \e \left(
  M + \s_2 + i \frac{q}{2f} \right) \Phi\ ,
  \label{missing}
\end{align}
is guaranteed to be a scalar eigenmode for $\D_b \doteq - M ( M+2\s_2 )$ via the map
$\bar \e \Psi_1$ (\ref{map0}). Hence, a spinor eigenmode for $\D_f \doteq M$ is missing, which results in the $\frac{1}{M}$ factor for the one-loop determinant.
The equation (\ref{missing}) can be managed into the following two equations
%%
%\begin{align}
%  -i \sin\frac{\th}{2} \left( \frac{1}{f} D_\th  - \frac{i}{l \sin\th} D_\varphi \right) \Phi
%  - i\cos{\frac{\th}{2}} \s_1  \Phi &
%  = + \cos\frac{\th}{2} \left( M + \s + i \frac{q}{2f} \right)  \Phi\ ,
%  \nonumber \\
%  -i \cos\frac{\th}{2} \left( \frac{1}{f} D_\th  + \frac{i}{l \sin\th} D_\varphi \right) \Phi
%  - i\sin{\frac{\th}{2}} \s_1  \Phi &
%  = -\sin\frac{\th}{2} \left( M + \s + i \frac{q}{2f} \right)  \Phi\ ,
%\end{align}
%%
%which can be recast as the following two equations
%
\begin{align}
  \partial_\varphi \Phi  = - \left( M l+ \s l  + i \frac{q}{2}  - i \frac{B\k}{2} \right) \Phi \ ,
  \label{mspinor1}
\end{align}
and
\begin{align}
  \frac{-i}{f} \partial_\th \Phi - \frac{\cos\th }{\sin\th} \left( M + \s + i \frac{q}{2f} \right) \Phi
  - \frac{i}{\sin\th} \s_1 \Phi = 0 \ .
  \label{mspinor2}
\end{align}
Assuming $\Phi(\th,\varphi) = e^{+i J \varphi} \chi(\th)$, the first equation (\ref{mspinor1}) determines
the eigenvalue $M$,
\begin{align}
  i J = - \left( M l  + \s l  + i \frac{q}{2} - i \frac{B\k}{2} \right) \ ,
\end{align}
while the second equation (\ref{mspinor2}) determines the unknown function $\chi(\th)$,
\begin{align}
  \frac{-i}{f} \partial_\th \chi + \frac{\cos\th }{\sin\th} \left( i \frac{J}{l} + i \frac{q}{2l} - i \frac{q}{2f} - i \frac{B\k}{2l} \right) \chi
  + \frac{i}{\sin\th} \frac{B}{2l}  \chi = 0 \ .
\end{align}
Again, it is enough to know the normalizability of the solutions. Looking at the differential equation, the solution can develop singularities at the north pole $\th=0$ and the south pole $\th=\pi$.
One can easily show that the solutions near the pole are approximate to
\begin{align}
  \chi(\th) \sim \sin^J\th \ ,
\end{align}
For the normalizability, one needs to require $J$ to be non-negative. Consequently,
the eigenvalues for unpaired spinors eigenmodes are given by
\begin{align}
  M l = - i \left( J - i\s_2 l + \frac q2 + \frac{|B|}{2} \right) \ ,
  \label{result2}
\end{align}
where $J \geq 0$.

\paragraph{one-loop determinant} For a given superselection sector $B$,
collecting all the results (\ref{result1}) and (\ref{result2}) results in
\begin{align}
  \left. \frac{\text{det}\D_f}{\text{det}\D_b} \right|_{B} & \simeq  \left. \frac{\text{det}\g^3\D_f}{\text{det}\D_b} \right|_{B}
  \nonumber \\ & \simeq
  \prod_{J=0}^\infty
  \frac{J +1 + i \s l -\frac q2 + \frac{|B|}{2}}{
  J - i\s l + \frac q2 + \frac{|B|}{2}}  \ ,
\end{align}
where the symbol `$\simeq$' represents the equality up to a sign factor
independent of the vacuum expectation value $\s$ but dependent on the flux $B$.
One can fix this factor from the comparison to the result for the round-sphere \cite{Doroud:2012xw,Benini:2012ui}.
That is to say, the sign factor is determined by $(-1)^{\frac{|B|+B}{2}}$. This result is exactly
the same to the result for the round two-sphere.

\subsection{Vector Multiplet}

Let us now in turn consider the one-loop determinant from the vector multiplet.
We denote the fluctuation modes for the vector and scalar fields as follows
\begin{align}
  A = \frac{B}{2} \left( \k - \cos\th \right) d\varphi + a \ , \qquad
  \s_1 = - \frac{B}{2l} + \z \ , \qquad
  \s_2 = \s + \eta \ .
\end{align}
With the Cartan-Wyel basis ($E_{\pm\a}, H$) where $\a$ denote the positive roots of G,
all the adjoint fields $\varphi$ can be decomposed as
\begin{align}
  \varphi = \sum_{i=1}^{r} \varphi^i H_i + \sum_{\a\in \D^+}
  \left( \varphi^\a E_{\a} + \varphi^{-\a} E_{-\a} \right) \ .
\end{align}
The contribution from the modes $\varphi^i$ to the one-loop determinant is $\sigma$-independent
and one can therefore ignore them in the discussion below.
One can show that the Laplacian operators action on bosonic fluctuations $(a^\a, \eta^\a, \z^\a)$ takes the following form
\begin{align}
  \D =
  \begin{pmatrix}
    - \ast D \ast D + (\a\cdot \s)^2 + \frac{\fq^2}{l^2} & - i (\a\cdot \s) D & i \frac{\fq}{l} D
    - \ast D  \frac1f
    \\
    - i (\a\cdot \s) \ast D \ast & - \ast D \ast D + \frac{\fq^2}{l^2} & (\a\cdot \s) \frac{\fq}{l}
    \\
    i \frac{\fq}{l} \ast D \ast + \frac1f \ast D & (\a\cdot\s) \frac{\fq}{l} & - \ast D \ast D
    + (\a\cdot\s)^2 + \frac{1}{f^2}
  \end{pmatrix}\ ,
\end{align}
where $\fq = \frac{\a\cdot B}{2}$ for $\a \in \D$.

\paragraph{non-physical modes}
One can easily find two eigenvectors $(a^\a, \eta^\a, \z^\a)$
of the operator $\D$,
\begin{align}
  \D \doteq 0 : & \ \ ( i D \eta , \ (\a\cdot\s) \eta , \ -\frac{\fq}{l} \eta )
  \nonumber \\
  \D \doteq \D_0 + (\a\cdot\s)^2 : & \ \
  ( - i (\a\cdot \s) D \eta ,\  \D_0 \eta ,\  \frac{\fq}{l} (\a\cdot \s) \eta )\ ,
  \label{nonphysical}
\end{align}
where $\D_0$ denotes an eigenvalue of an operator `$-\ast D \ast D + \frac{\fq^2}{l^2}$'. Obviously, the zero eigenmodes of $\D$ corresponds to
gauge symmetry which should not be regarded as physical modes.
In order to evaluate the path integral properly,
one needs to take into account for the Faddeev-Popov determinant.
We however choose a rather short-cut discussed in \cite{Hama:2011ea},
instead of considering the ghost fields and
some modifications on supersymmetry by BRST transformtaion. One can show that
the contribution from another eigenmodes in (\ref{nonphysical}) for $\D \doteq \D_0 + (\a\cdot\s)^2$ is exactly canceled by the Faddeev-Popov determinant,
\begin{align}
  1 = \CJ \cdot \int D'w \ \text{Exp}\Big[ -\frac12 \int \
  \text{tr}\left( w \wedge \ast \big(-\ast D\ast D  + \frac{\fq^2}{l^2} + (\a\cdot\s)^2 \big) w \right)\Big]\ ,
\end{align}
which arises from the gauge fixing excluding the contribution from zero eigenvalues.
For details, please refer to \cite{Hama:2011ea}.

\paragraph{super multiplet} The four bosonic eigenmodes for ($a^\a, \eta^\a, \z^\a$) split
into two covariant longitudinal-modes with $a^\a \sim D\eta^\a$ and two transverse modes with
$a^\a \sim \ast D \eta^\a$. As explained above,
the covariant longitudinal-modes results in no net contribution to the one-loop determinant.
The problem is therefore how two phsyical bosonic-modes can be paired with
the spinor eigenmdes.

Fixing a gauge\footnote{In order to set $\eta^\a=0$, the vector field $a^\a$
has a covariant longitudinal piece. However, this gauge is rather convenient to see how
bosonic and spinor modes are paired.},
\begin{align}
  \ast D \ast a^\a = -i \frac{\fq}{l} \z^\a \ , \qquad
  \eta^\a = 0\ ,
  \label{gaugechoice}
\end{align}
the differential operators of our interest under the gauge choice (\ref{gaugechoice})
become
\begin{align}
  \D_b & =
  \begin{pmatrix}
  - \ast D \ast D + (\a \cdot \s)^2 + \frac{\fq^2}{l^2} &
  i \frac{\fq}{l} D - \ast D \frac 1f
  \\
  i \frac{\fq}{l} \ast D \ast  + \frac{1}{f} \ast D & - \ast D \ast D + \frac{1}{f^2} + (\a\cdot \s)^2
  \end{pmatrix}  \ ,
  \nonumber \\
  \D_f & = i \g^i D_i - i \frac{ \fq}{l}  + (\a\cdot \s) \g^3\ ,
\end{align}
where the operator $\D_b$ acts on $(a^\a, \z^\a)$.
For later convenience, it is useful to consider the eigenmodes for the following operators,
\begin{align}
  \d_b \equiv \begin{pmatrix} i (\a\cdot \s) + i \frac{\fq}{l} \ast & - \ast D \\
  \ast D & \frac{1}{f} + i (\a\cdot \s) \end{pmatrix} \ , \qquad
  \g^3 \D_f \ .
\end{align}
One can show that, for any eigenmode of $\d_b\doteq - i M$,
\begin{align}
  \d_b \d_b = \D_b + 2i (\a\cdot\s) \d_b \ \to \
  \D_b \doteq - M \big( M + 2(\a\cdot\s) \big)\ .
\end{align}
It implies that
\begin{align}
  \det \D_b = \left( \det \d_b \right)^2\ .
\end{align}

For the vector multiplet, the bosonic and sermonic eigenmodes can be paired
in the following manner. Let $(\CA,\S)$ denote a vector/scalar eigenmode for
$\d_b \doteq -iM$. A spinor eigenmode for $\g^3\D_f \doteq - M$ can be
obtained from the map below
\begin{align}
  \L = \left( \g^3\g^i \CA_i + i \S \g^3 \right) \e \ .
\end{align}
On the other hand, the map from a spinor eigenmode for $\g^3\D_f \doteq -M$
to a vector/scalar eigenmode for $\d_b\doteq - i M $ is
\begin{gather}
  \CA = - i \left( M + \a\cdot\s \right) \bar \e \g_i \L e^i
  - D \left( \bar \e \g^3 \L \right)\ ,
  \nonumber \\
  \S = \left( M + \a\cdot \s \right) \bar \e \L
  - i \frac{\fq}{l} \bar \e \g^3 \L\ .
  \label{map4}
\end{gather}
The relation could be guessed from the supersymmetry transformation
laws (\ref{SUSYvec}) with some care given to the gauge condition (\ref{gaugechoice}).
Any paired modes via the above maps are irrelevant when computing the one-loop determinant. As in the case for the matter multiplet, one can have nontrivial contribution to the one-loop determinant from unpaired and
missing spinor eigenmodes.

\paragraph{unpaired spinor eigenmodes} Unpaired spinor eigenmodes are annihilated
by the map (\ref{map4}): one can first show that
\begin{align}
  \S = 0 \ \to \ \L = g(\th,\varphi) \left( \bar \e + \frac{i \fq}{ l \left(
  M + \a\cdot\s\right)} \g^3\bar \e \right) \ .
  \label{ansatz4}
\end{align}
Using the above expression for $\L$, one needs to satisfy the condition
$\CA=0$ and the spinor eigenmode equation $\g^3\D_f \L = - M \L$ simultaneously.
At first, it looks overdetermined but it turns out that one relation implies another.
One can show that the undetermined function $g(\th,\varphi)$ should satisfy
\begin{align}
  \partial_\varphi g - i g - i \fq \k g + l \left( M +\a\cdot\s \right) g = 0\ ,
  \label{eqn5}
\end{align}
and
\begin{align}
  \frac1f \sin \th \partial_\th g + \cos \th \left( \frac 1f + i
  \left( M + \a\cdot\s \right) \right) g + \frac\fq{l} g = 0 \ .
  \label{eqn6}
\end{align}
Using the ansatz $g(\th,\varphi) = e^{- i J \varphi} h(\th)$,
the first equation (\ref{eqn5}) determines the eigenvalues $M$
\begin{align}
  M l  = + i \left( J +1 + \fq \k + i  \a\cdot \s l \right)\ .
\end{align}
Solving the second equation (\ref{eqn6}) determines the function $h(\th)$
\begin{align}
  \frac1f \sin \th \partial_\th h + \cos \th \left( \frac 1f
   - \frac{J + 1 + \fq \k }{l} \right) h + \frac\fq{l} h = 0 \ .
\end{align}
It is enough to know when the solution becomes normalizable.
Near the poles $\th=0,\pi$, the solutions are approximate to
\begin{align}
  h(\th) \sim \sin^J \th \ .
\end{align}
Naively, the normalizablity could imply that $J$ is non-negative.
However, with a careful look at (\ref{ansatz4}), one can show that
the eigenmode of $J=-1$, i.e., $l \left( M + \a\cdot \s \right) = i |\bf q|$,
are indeed normalizable. We have the following cases:
\begin{enumerate}

  \item for the positive $\fq>0$, the spinor eigenmode (\ref{ansatz4}) of $J=-1$  (or equivalently $\k=1$),
  would-be-singular at the north pole,  becomes
  \begin{align}
    \L \sim \begin{pmatrix} \frac{1}{\cos\frac{\th}{2}}\\ 0 \end{pmatrix}\ ,
  \end{align}
  which is smooth at the north pole $\th = 0$.

  \item for the negative $\fq<0$, the spinor eigenmode
  (\ref{ansatz4}) of $J=-1$ (or equivalently $\k=-1$), would-be-singular at the south pole, becomes
  \begin{align}
    \L \sim \begin{pmatrix} 0 \\ \frac{1}{\sin\frac{\th}{2}} \end{pmatrix}\ ,
  \end{align}
  which is smooth at the south pole $\th=\pi$.

  \item One remark is that such modes of $J=-1$ become non-normalizable
  when $B =0$.
\end{enumerate}
As a summary, the eigenvalues for unpaired spinor eigenmodes are given by
\begin{align}
  M l & = + i \Big( J + |\fq| + i  \a\cdot \s l \Big)
  \ \  \text{ for } B \neq 0\ ,
  \nonumber \\
  M l & = + i \Big( J + 1 + i  \a\cdot \s l \Big)
  \ \ \ \ \text{ for } B = 0 \ ,
\end{align}
where $J \geq 0$.

\paragraph{missing spinor eigenmodes} In order to work out
the missing spinor eigenmodes,
one begins by solving $\CA_i \g^3 \g^i \e + i \S \g^3\e = 0$. It leads to
\begin{align}
  \CA= g(\th,\varphi) \left( e^1 + i \cos\th e^2 \right) \ , \ \
  \S = i g(\th,\varphi) \sin\th\ .
  \label{ansatz5}
\end{align}
Using the above expression for $(\CA,\S)$, one needs to satisfy the
eigenmode equations $\d_b \doteq - i M$ and the gauge condition
(\ref{gaugechoice}) simultaneously.
At first, it again looks overdetermined but turns out that any
solutions to one of the eigenmode equation with the ansatz (\ref{ansatz5}),
\begin{align}
  \ast D \CA + \left( \frac1f + i ( M + \a\cdot \s ) \right) \S = 0\ ,
\end{align}
automatically satisfy the other relations.
That is to say, it is sufficient to solve
the following two equations
\begin{align}
  \partial_\varphi g + \Big( l (M + \a\cdot\s) - i \fq \k \Big) g = 0 \ ,
  \label{eqn7}
\end{align}
and
\begin{align}
  \frac1f \Big( \sin\th \partial_\th  + \cos \th  \Big) g - \frac\fq{l} g
  - i \cos\th \left( M + \a\cdot \s \right) g = 0 \ .
  \label{eqn8}
\end{align}
Putting the ansatz $g(\th,\varphi) = e^{+ i (J+1) \varphi} h(\th)$, the first equation
(\ref{eqn7}) determines the eigenvalue $M$,
\begin{align}
  M l  = - i \Big( J + 1 - \fq \k - i \a\cdot \s l \Big)\ .
\end{align}
The second equation which determines the function $h(\th)$ can be
rewritten as
\begin{align}
  \frac1f \sin\th \partial_\th h + \frac1f \cos\th h -
  \frac\fq{l} h - \cos\th \left( \frac{J+1-\fq\k}{l} \right) h = 0\ .
\end{align}
Again, one can show that near the poles $\th=0,\pi$ the solutions are approximate to
\begin{align}
  h(\th) \sim \sin^J\th\ .
\end{align}
For the normalizablility, one needs to require $J$ to be non-negative. As a summary,
the eigenvalue for missing spinor eigenmodes are given by
\begin{align}
  M l = - i \Big( J +1 + |\fq| - i \a\cdot\s l \Big)\ ,
\end{align}
where $J \geq 0$.

\parskip 0.2cm
\paragraph{one-loop determinant} For a given superselection sector $B\neq 0$, the
contribution to the one-loop determinant from the vector multiplet results in
\begin{align}
  \left. \frac{\det \D_f}{\sqrt{\det \D_b}} \right|_{B\neq 0}
  & \simeq \left. \frac{\det \g^3 \D_f}{\det \d_b} \right|_{B\neq 0}
  \nonumber \\ & \simeq \prod_{\a \in \D_\text{root}} \prod_{J=0}
  \frac{J  + \left|\frac{\a\cdot B}{2} \right| - i (\a\cdot\s l )  }{J + 1 + \left|\frac{\a\cdot B}{2} \right| + i (\a\cdot\s l)  }
  \nonumber \\ & \simeq \prod_{\a \in \D_+} \Bigg[
  (\a\cdot \s l )^2 + \left( \frac{\a\cdot B}{2} \right)^2 \Bigg]\ .
\end{align}
By comparing the above results to those in the round-sphere,
one can fix a $\s-$independent factor by the unity. In the sector of $B =0$,
one gets the trivial result for the one-loop determinant. The result is
the same to the result for the round two-sphere \cite{Doroud:2012xw,Benini:2012ui}.
\vfill\eject
\bibliography{refs}

\end{document}